\documentclass[a4paper,american,aps,citeautoscript,floatfix,longbibliography,pdftex,pre,superscriptaddress,showpacs,preprint]{revtex4-1}
\usepackage{amssymb, graphicx,subfigure}
\usepackage{enumerate,tensor}
\usepackage{amsmath,bm}
\usepackage{dcolumn}
\usepackage{color}
\usepackage{natbib} 
\usepackage[latin1]{inputenc}
\usepackage{subfigure}
\usepackage{textcomp}
\usepackage{hyperref}
\usepackage{booktabs}
\usepackage{makecell}
\usepackage{psfrag}
\usepackage{epsfig}
\usepackage{ulem}
\newcommand{\w}{\ensuremath{\omega}}%

\makeatletter
\def\paragraph{\@startsection{paragraph}{4}{10pt}{-1.25ex plus -1ex minus -.1ex}{0ex plus 0ex}{\normalsize\textit}}
\renewcommand\@biblabel[1]{#1}
\renewcommand\@makefntext[1]%
{\noindent\makebox[0pt][r]{\@thefnmark\,}#1}
\DeclareRobustCommand\onlinecite{\@onlinecite}
\def\@onlinecite#1{\begingroup\let\@cite\NAT@citenum\citealp{#1}\endgroup}
\def\tagform@#1{\maketag@@@{\ignorespaces#1\unskip\@@italiccorr}}
\let\orgtheequation\theequation
\def\theequation{(\orgtheequation)}
\makeatother

\begin{document}

\title{Equilibria and Dynamics of two coupled chains of interacting  dipoles}

\author{Manuel I\~narrea}
\affiliation{\'Area de F\'{\i}sica, Universidad de La Rioja, 26006 Logro\~no, La Rioja, Spain}

\author{Rosario Gonz\'alez-F\'erez}
\affiliation{Instituto Carlos I de F\'{\i}sica Te\'orica y Computacional,
and Departamento de F\'{\i}sica At\'omica, Molecular y Nuclear,
  Universidad de Granada, 18071 Granada, Spain}
  
\author{J. Pablo Salas}
\affiliation{\'Area de F\'{\i}sica, Universidad de La Rioja, 26006 Logro\~no, La Rioja, Spain}

\author{Peter Schmelcher}
\affiliation{The Hamburg Center for Ultrafast Imaging, Luruper Chaussee 149, 22761 Hamburg, Germany}
\affiliation{Zentrum f\"ur Optische Quantentechnologien, Universit\"at
  Hamburg, Luruper Chaussee 149, 22761 Hamburg, Germany} 

\date{\today}
\begin{abstract} 
We explore the energy transfer dynamics in an array of two chains of identical rigid interacting dipoles. A crossover between two different ground state (GS) equilibrium configurations is observed with varying distance between the two chains of the array. Linearizing around the GS configurations, we verify that interactions up to third nearest neighbors
 should be accounted for accurately describe the resulting dynamics. Starting with one of the GS, we excite the system by supplying it with an excess energy $\Delta K$ located initially on one of the dipoles. We study the time evolution of the array for different values of the system parameters $b$ and $\Delta K$. Our focus is hereby on two features of the energy propagation: the redistribution of the excess energy $\Delta K$ among the two chains and the energy localization along each chain. For typical parameter values, the array of dipoles reaches both the equipartition between the chains and the thermal equilibrium from the early stages of the time evolution. Nevertheless, there is a region in parameter space $(b,\Delta K)$ where even up to the long computation time of this study, the array does neither reach energy equipartition nor thermalization between chains. This fact is due to the existence of persistent chaotic breathers.
\end{abstract}

\maketitle

\section{Introduction}
The intriguing results obtained by Fermi, Pasta, Ulam, and Tsingou (FPUT) in 1953 \cite{fput,A780,today} in their study of the energy relaxation in a chain of nonlinear oscillators have been driving much of the research that, since then, has been carried out on numerous nonlinear Hamiltonian lattices.
Much of the great interest in nonlinear lattices lies in the fact that they show collective effects,  behaviors that are in no way characteristic of systems with few degrees of freedom. For example, the spontaneous appearance of chaotic breather-like excitations in Hamiltonian lattices is a collective effect that plays a fundamental role in energy transfer processes because they may cause the thermalization of the system to be extremely slow. Different nonlinear systems exhibit this behavior such as lattices of FPUT-like oscillators, Klein-Gordon oscillators, Josephson junctions, Bose-Einstein condensates, Heisenberg spins or rigid electric dipoles\cite{A908,A1000,A998,A997,A1147,A1148,chain}.

Although many examples of studies on two-dimensional lattices with different kinds of oscillators can be found \cite{A922,A921,A915,A872,A1256,A923,A1103,A1116,A920,A1251,A1253,A1254,A1255,A1252}, a common denominator in the vast literature on nonlinear lattices is that, in general, most of the studies are reduced to one-dimensional systems, i.e.,  to linear chains with different boundary conditions. The investigations in nonlinear lattices in two or three dimensions are more scarce and less developed.
Furthermore, in most cases, and regardless of the dimension of the lattice, the oscillators are coupled via interactions that are usually only extended up to nearest neighbors. However, in the case of long-range interactions, the nearest neighbors approach may not be well-justified.

Motivated by the above, we address here lattices beyond nearest neighbors interactions. This latter condition is easily met when the lattice is made up of interacting dipoles.
One implementation here are cold polar diatomic molecules trapped in optical lattices that exhibit an intriguing quantum collective dynamics \cite{A753,rey,Lewenstein}.
In this way, the classical approach of considering trapped cold (but not ultracold) polar molecules as linear chains of interacting rigid dipoles has been used by several authors
\cite{Ratner1,Ratner2,Ratner3,zampetaki} to study the energy transfer in various planar configurations.
Indeed, the energy transfer has been shown to lead to the formation of solitons
or to the emergence of chaoticity \cite{zampetaki,A898}.
We note that even the simplest two-dipole chain \cite{PRE2017,CNSNS2020} was found to display a rich dynamical behavior. More recently, the connection between chaos, thermalization and ergodicity has been study \cite{chain}.

On the other hand, an immediate extension of a one-dimensional lattice would be a one-dimensional array made of a limited number of parallel linear chains. In the particular case of an array of two linear chains, particles can be arranged according to two main configurations, namely in a ladder array or in a sawtooth array. Note that, besides linear chains of oscillators with alternating masses \cite{solidsbook,A1187} or with alternating kind of interactions \cite{A1176}, one dimensional arrays are the simplest lattices showing a multiband structure. An example can be found in Ref.\cite{A1021} where the dynamical properties of a sawtooth array of frustrated Josephson junctions has been studied.

In this work we perform a dynamics study of a ladder array formed by two chains of dipoles.  Arrays of dipoles can be prepared experimentally by trapping cold or
ultracold dipolar molecules in optical lattices \cite{Kotochigova,Capogrosso,Schachenmayer} where the wavelength
of the light sets the scale for the interaction strength among neighboring molecules.
In particular, an experimental realization of an array of two chains can be obtained by employing superlattices of double wells \cite{Neustetter} in combination
with a regular optical lattice for the different spatial directions.
Other possible candidates for an experimental implementation are
colloidal polar particles in optical tweezers \cite{Mittal}.

In this study we focus on two main goals.
On the one side, we investigate the impact of dipolar interactions beyond nearest neighbors: We determine up to what order the interaction should be taken into account such that the dynamics of the system is properly described. On the other side, we explore the different energy transfer mechanisms that the ladder array of dipoles shows when it is subjected to single site excitations.

This work is organized as follows. In Sec. II we provide the Hamiltonian that governs the dynamics of the system. The ground state (GS) configurations of the system are analyzed in Sec.II: Depending on the separation between the two chains we observe two different GS configurations. In Secs. III and IV we study the linearized dynamics of the system around those equilibria demonstrating that it is not affected by the inclusion of interaction ranges beyond third neighbors.
Starting form the GS configuration we follow in Sec. V the time evolution of an initial single site excitation. In particular, for different excitation energies, for different values of the separation between the two chains, and for different time windows, we compute the time-average of the total energy located in each chain of the array, as well as the corresponding time-averages of the participation function \cite{A918,chain}. This later calculation provides information about the degree of thermalization of each chain. Our conclusions are provided in Sec.VI.

\begin{figure}[h]
\centerline{\includegraphics[scale=0.5]{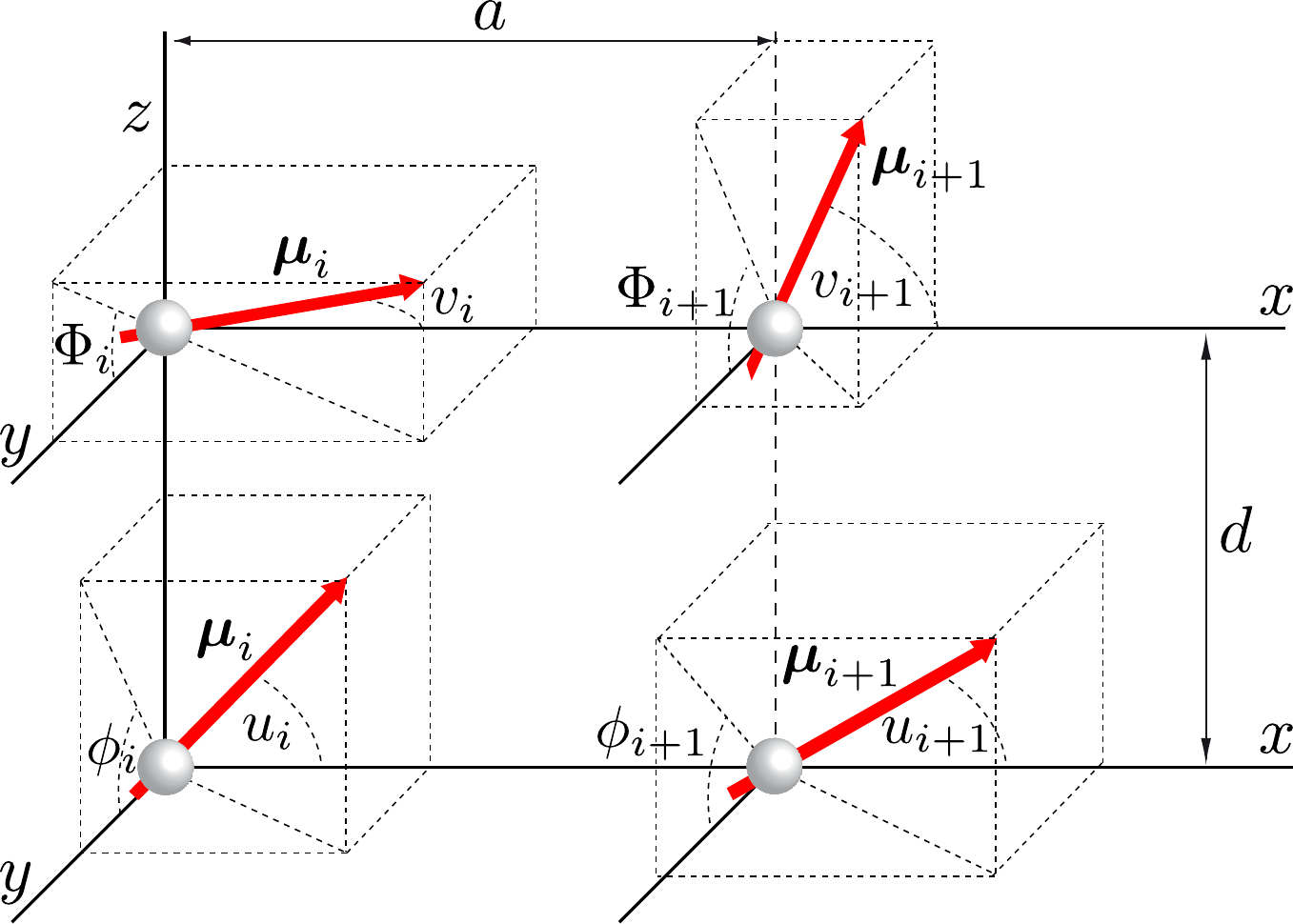}}
\caption{Schematic representation of the ladder array of two dipole chains.}
\label{fi:Euler}
\end{figure}

\section{Hamiltonian and equilibrium configurations}
\label{sec:II}
The potential energy $V_{i j}$ between two dipoles with dipole moments $\boldsymbol{\mu}_i$ and $\boldsymbol{\mu}_j$ is given by
\begin{equation}
\label{dipole}
V_{i j}=\frac{1}{4 \pi \epsilon_0} \frac{(\boldsymbol{\mu}_i \cdot \boldsymbol{\mu}_j) r_{i j}^2-3 (\boldsymbol{\mu}_i \cdot {\bf r}_{i j})(\boldsymbol{\mu}_j \cdot {\bf r}_{i j})}{r_{i j}^5},
\end{equation}
where ${\bf r}_{i j}$ is the relative vector of the positions of the dipoles $i$ and $j$. 
We consider a ladder array of two chains of $N$ identical dipoles. According to Fig.\ref{fi:Euler}, the $N$ dipoles of each chain are fixed in space along the $x$-axis of the Laboratory Fixed Frame $xyz$ with a distance $a$ between two consecutive dipoles. The distance between the chains is $d$.
Using Euler angles (see Fig.\ref{fi:Euler}), the dipole moments of the rotors belonging to the
lower and the upper chains are given by the vectors
\begin{eqnarray}
\nonumber
&&\{\mu \ (\cos u_1,\cos \phi_1\sin u_1,\sin u_1\sin \phi_1), ..., \mu \ (\cos u_N,\cos \phi_N\sin u_N,\sin u_N\sin \phi_N),\\
 \label{moments}
&& \mu \ (\cos v_1,\cos \Phi_1\sin v_1,\sin v_1\sin \Phi_1), ..., \mu \ (\cos v_N,\cos \Phi_N\sin v_N,\sin v_N\sin \Phi_N)\},
\end{eqnarray}
where $0 \le (u_i, v_i) < \pi$, and  $0 \le (\phi_i, \Phi_i) < 2\pi$ (see Fig.\ref{fi:Euler}).

\begin{figure}[h]
\centerline{\includegraphics[scale=0.4]{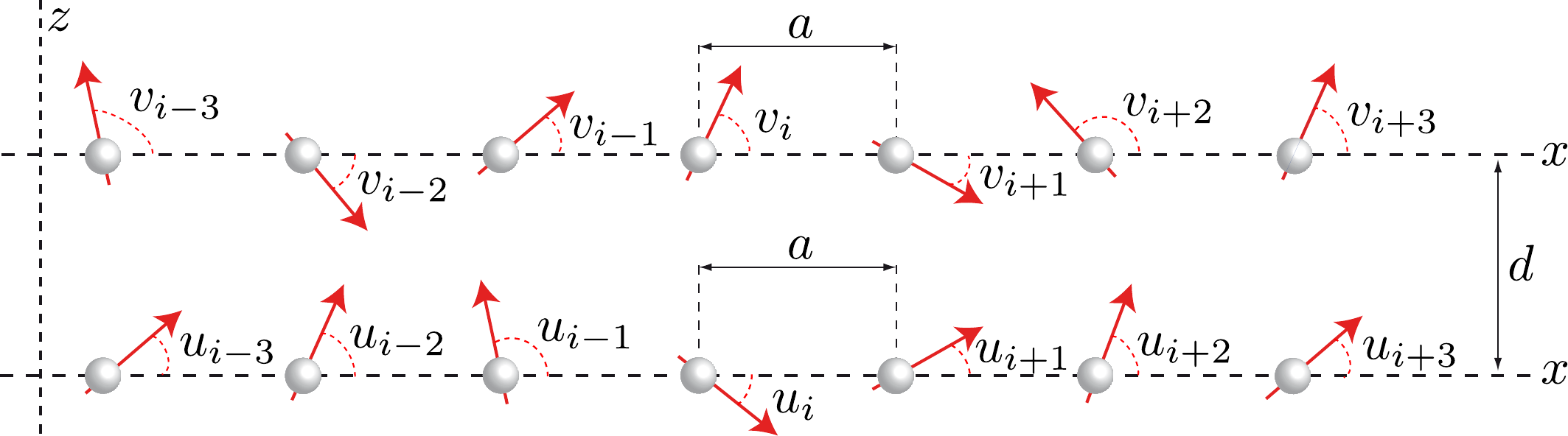}}
\caption{Schematic representation of the ladder array of two dipole chains in the invariant manifold ${\cal M}$.}
\label{fi:chain1}
\end{figure}

The total interaction potential $V$ is made of three main terms. The interactions $V_1(u_i,\phi_i,u_j,\phi_j)$ between the dipoles belonging to the lower chain, the interactions $V_2(v_i,\Phi_i,v_j,\Phi_j)$ between the dipoles belonging to the upper chain, and the interactions $V_3(u_i,\phi_i,v_j,\Phi_j)$ between the dipoles of the lower chain and the dipoles of the upper chain. Using the general term \ref{dipole} and the expressions of the dipole moments \ref{moments}, these three terms are
\begin{eqnarray}
\label{term1}
V_1(u_i,\phi_i,u_j,\phi_j)&=&\frac{\mu^2}{a^3 (i-j)^3}(\sin u_i \sin u_j \cos(\phi_i-\phi_j)-2\cos u_i \cos u_j),\\[2ex]
\label{term2}
V_2(v_i,\Phi_i,v_j,\Phi_j)&=&\frac{\mu^2}{a^3 (i-j)^3}(\sin v_i \sin v_j \cos(\Phi_i-\Phi_j)-2\cos v_i \cos v_j),\\[2ex]
\nonumber
V_3(u_i,\phi_i,v_j,\Phi_j)&=&\frac{\mu^2}{(d^2+a^2 (i-j)^2)^{5/2}}((d^2-2 a^2(i-j)^2)\cos u_i \cos v_j+\\[2ex]
\label{term3}
&& 3 a d (i-j) \cos u_i\sin v_j \sin \Phi_j+3 a d (i-j) \sin u_i\cos v_j \sin \phi_j+\\[2ex]
\nonumber
&& (d^2+a^2(i-j)^2)\sin u_i \sin v_j \cos\phi_i \cos \Phi_j+\\[2ex]
\nonumber
&& (a^2(i-j)^2-2 d^2)\sin u_i \sin v_j \sin\phi_i \sin \Phi_j).
\nonumber
\end{eqnarray}
The rotational dynamics of the system, as a function of the phases ${\bf x}=\{(u_i, \phi_i, v_i, \Phi_i), \ i=1, ..., N\}$, is formally described by the following Hamiltonian (corresponding to the energy $E$)
\begin{equation}
\label{hamiEuler1}
{\cal H}\equiv E=\frac{1}{2 I}\sum_{i=1}^{N}\left(P_{u_i}^2+\frac{P_{\phi_i}^2}{\sin u_i^2}+
P_{v_i}^2+\frac{P_{\Phi_i}^2}{\sin v_i^2}\right) + V(u_i, \phi_i, v_i, \Phi_i), 
\end{equation}
where $I$ is the moment of inertia of the dipoles.
The term $V(u_i, \phi_i, v_i, \Phi_i)$ is the total interaction potential of the system given by
\begin{equation}
V(u_i, \phi_i, v_i, \Phi_i)=\sum_{i<j}^N \left(V_1(u_i,\phi_i,u_j,\phi_j)+V_2(v_i,\Phi_i,v_j,\Phi_j)\right)+\sum_{i,j}^N V_3(u_i,\phi_i,v_j,\Phi_j).
\end{equation}
The Hamiltonian \ref{hamiEuler1} defines a dynamical system with 4$N$ degrees of freedom, where $P_{u_i}$, $P_{\phi_i}$, $P_{v_i}$ and $P_{\Phi_i}$ are the conjugate momenta of $u_i$, $\phi_i$, $v_i$ and $\Phi_i$ respectively.
From the inspection of Hamiltonian \ref{hamiEuler1}, we see that the manifold ${\cal M}$ of codimension 2$N$ given by
\begin{equation}
{\cal M} = \lbrace(u_i, P_{u_i}, v_i, P_{v_i}) \ | \ \phi_i=\Phi_i=\pi/2 \ \mbox{and} \ P_{\phi_i}=P_{\Phi_i}=0\rbrace,
\end{equation}
is invariant under the dynamics, such that the number of
degrees of freedom of the system is reduced to $2N$.
On the manifold ${\cal M}$, the Hamiltonian \ref{hamiEuler1} reads
\begin{equation}
\label{hamiEulerM}
{\cal H}=\frac{1}{2 I}\sum_{i=1}^{N}\left(P_i^2+
Q_i^2\right) + V(u_1,...,u_N,v_1,...,v_N), 
\end{equation}
such that the momenta $(Q_i, P_i)=(P_{u_i}, P_{v_i})$ are $Q_i=I \ d u_i/dt$ and $P_i=I \ d v_i/dt$.
We suspect that this invariant subspace ${\cal M}$ represents a stable
manifold. Irrespective of this, a standard way of suppressing additional
dimensions in e.g. ultracold atomic gases is to provide a strongly
confining typically harmonic potential along the transversal degrees of
freedom which are in our case the $y$-coordinates of the dipoles. The
latter can be achieved by e.g. optical trapping and standing light
waves. This way our Hamiltonian could be seen as an effective or reduced
Hamiltonian having traced out the fast transversal degrees of freedom.
From now on, we focus on the planar dynamics arising from
the Hamiltonian \ref{hamiEulerM}, i.e., we assume that dipoles are restricted to rotate in the common $xz$-plane (see Fig.\ref{fi:chain1}).

One of the goals of this article is to explore the impact of interactions beyond nearest neighbors on the dynamics of the dipoles. Thereby, we will determine up to what order we should account for the interactions such that the dynamics is described accurately. The strategy will be to determine, in a linear approximation, the relationship between the rotational frequency and the wave number, i.e., the dispersion relation, such that the progressive inclusion of more distant neighbours in the interaction does not affect the dispersion relation significantly.

Periodic boundary conditions (pbc)  within each chain
are assumed. Up to a given $r \ge 1$ interaction order, the potential $V=V(u_1,..., u_N, v_1,..., v_N)$ in Hamiltonian \ref{hamiEulerM} can be written as
\begin{equation}
\label{potential1}
V=\chi \sum_{n=1}^N\left(\sum_{j=1}^r [V_1(u_n,u_{n+j}) + V_2(v_n,v_{n+j})]+\sum_{k=-r}^r V_3(u_n,v_{n+k})\right),
\end{equation}
where $\chi=\mu^2/4 \pi \epsilon_0 a^3$ is a parameter that measures the strength of the dipole-dipole interaction. The terms $V_1(u_n,u_{n+j})$, $V_2(v_n,v_{n+j})$ and $V_3(u_n,v_{n+j})$ in Eq.\ref{potential1}are
\begin{eqnarray}
\label{v1}
V_1(u_n,u_{n+j})&=&\frac{\left(\sin u_n \sin u_{n+j} - 2 \cos u_n \cos u_{n+j} \right)}{j^3},\\[2ex]
\label{v2}
V_2(v_n,v_{n+j})&=&\frac{\left(\sin v_n \sin v_{n+j} - 2 \cos v_i \cos v_{n+j} \right)}{j^3},\\[2ex]
\label{v3}
V_3(u_n,v_{n+k})&=&\frac{(k^2-2 b^2) \sin u_n \sin v_{n+k}+(b^2-2 k^2) \cos u_n \cos v_{n+k}}{(b^2+k^2)^{5/2}}-\\[2ex]
\nonumber
&& \frac{3 b k \sin (u_n+v_{n+k})}{(b^2+k^2)^{5/2}}
\end{eqnarray}
where $b=d/a$ is the distance between the chains in units of $a$. The terms $V_1$ and $V_2$ describe the interactions between the dipoles of the lower and the upper chains, respectively. The term $V_3$ accounts for the interactions between dipoles belonging to different chains.

Defining ${\bf X}$ as the vector of the phase space variables
\[
{\bf X}=(u_1,..., u_N, v_1,..., v_N, Q_1,..., Q_N, P_1,..., P_N),
\]
the Hamiltonian equations of motion are obtained as $\dot{{\bf X}}={\cal L}_{{\cal H}} {\bf X}$
where 
\[
{\cal L}_{{\cal H}}=-\sum_{n=1}^{2N}\left(\frac{\partial {\cal H}}{\partial X_n} \frac{\partial }{\partial X_{n+2N}}-\frac{\partial {\cal H}}{\partial X_{n+2N}} \frac{\partial }{\partial X_n}\right).
\]

\medskip\noindent
The first step in describing the dynamics of our system is the determination of the
ground state (GS), which corresponds to the equilibrium point of  of Hamiltonian flow \ref{hamiEuler1}. When they exist, equilibria appear when $Q_n=P_n=0$. Thence, they actually correspond to the critical points of the potential energy surface $V$ given by Eq.\ref{potential1}. Therefore, the critical points of $V$ are the roots of the system of equations

\begin{eqnarray}
\label{partialU}
\frac{\partial V}{\partial u_n}&=&\chi \sum_{j=1}^r \frac{(\sin u_{n-j}+\sin u_{n+j})\cos u_n +2 (\cos u_{n-j}+
\cos u_{n+j}) \sin u_n }{j^3}+\\
\nonumber
&& \chi \sum_{k=-r}^r \frac{(k^2-2 b^2) \cos u_n \sin v_{n+k}-(b^2-2 k^2) \sin u_n \cos v_{n+k} -3 b k \cos (u_n+v_{n+k})}{(b^2+k^2)^{5/2}}=0,\\
\label{partialV}
\frac{\partial V}{\partial v_n}&=&\chi \sum_{j=1}^r \frac{(\sin v_{n-j}+\sin v_{n+j}) \cos v_n +2 (\cos v_{n-j}+
\cos v_{n+j}) \sin v_n }{j^3}+\\
\nonumber
&& \chi \sum_{k=-r}^r \frac{(k^2-2 b^2) \cos v_n \sin u_{n+k}-(b^2-2 k^2) \sin v_n \cos u_{n+k} +3 b k \cos (v_n+u_{n+k})}{(b^2+k^2)^{5/2}}=0.
\end{eqnarray}
\noindent
From the direct inspection of the equations \ref{partialU}-\ref{partialV}, we conclude that equilibria appear for the following general configurations $C_1$, $C_2$ and $C_3$:
\begin{itemize}
\item[i)] $C_1=\{u_n=v_n=0, \forall n\}$ or $C_1=\{u_n=v_n=\pi, \forall n\}$. The energy $E_1$ of this head-tail configuration is
\begin{equation}
\label{enerC1}
E_1=-4\chi N \sum_{j=1}^r \frac{1}{j^3}+\frac{\chi N}{b^3}+2\chi N \sum_{j=1}^r \frac{b^2-2 j^2}{(b^2+j^2)^{5/2}}.
\end{equation}
In the case of two non-interacting chains (i.e., $b\rightarrow \infty$), we recover
the expected value $E_1=-4\chi N \sum_{j=1}^r 1 /j^3$, which corresponds to the ground state of two separate chains of length $N$ each with p.b.c.

\item[ii)] $C_2=\{u_n=0, v_n=\pi, \forall n\}$ or $C_2=\{u_n=\pi, v_n=0, \forall n\}$. The energy $E_2$ of this head-tail configuration, where dipoles of each chain are oppositely oriented, is
\begin{equation}
\label{enerC2}
E_2=-4\chi N \sum_{j=1}^r \frac{1}{j^3}-\frac{\chi N}{b^3}-2\chi N \sum_{j=1}^r \frac{b^2-2 j^2}{(b^2+j^2)^{5/2}}.
\end{equation}
We note that $E_2<E_1$.

\item[iii)] $C_3=\{u_n=v_n=(-1)^n \pi/2, \forall n\}$ or $C_3=\{u_n=v_n=(-1)^{n+1} \pi/2, \forall n\}$. The energy $E_3$ of this head-tail
configuration between chains, where neighboring dipoles are in alternating orientations $\pm \pi/2$, is
\begin{equation}
\label{enerC3}
E_3=2\chi N \sum_{j=1}^r \frac{(-1)^j}{j^3}-\frac{\chi N}{b^3}+2 \chi N \sum_{j=1}^r (-1)^{j} \frac{j^2-2 b^2}{(b^2+j^2)^{5/2}}.
\end{equation}
\end{itemize}
\begin{figure}[t]
\centerline{\includegraphics[scale=0.6]{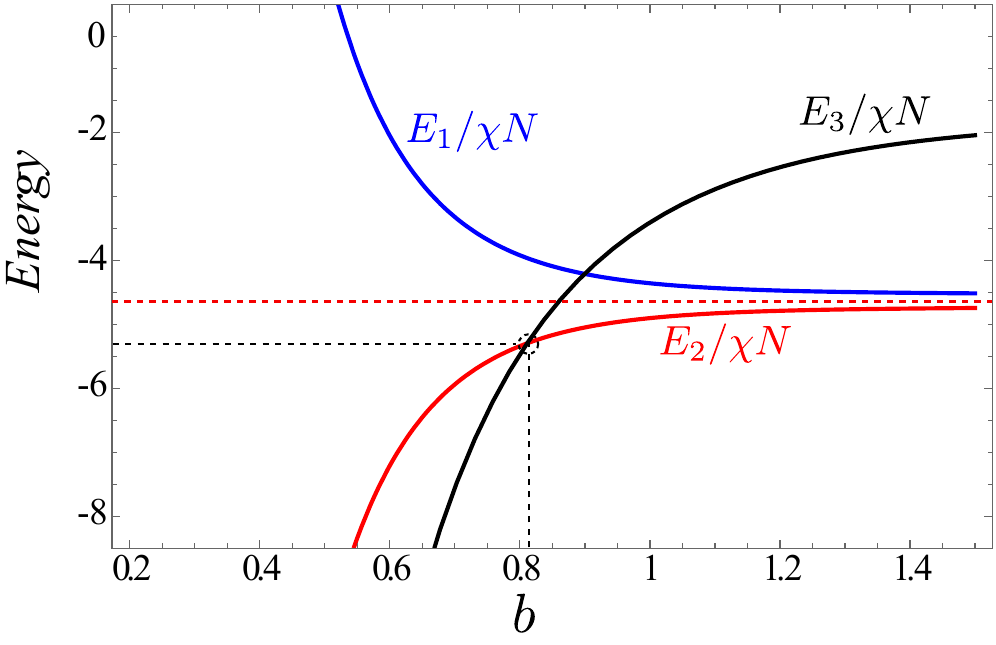}}
\caption{The evolution of the energies $E_1$ (blue line), $E_2$ (red line) and $E_3$ (black line) of the equilibrium configurations $C_1$, $C_2$ and $C_3$ as a function of the distance parameter $b$ up to an interaction of order $r=3$.
The energy is measured is units of $E_o=\chi N$. The dashed black lines and circle mark the energy crossing between the
energies $E_2$ and $E_3$.
The red dashed line indicates the asymptotic energy $E_{\infty}/\chi N=-4 \sum_{j=1}^r \frac{1}{j^3}$, which corresponds to the energy of two non-interacting chains of $N$ dipoles.}
\label{fi:energies}
\end{figure}

The energies $E_1$, $E_2$ and $E_3$ of the equilibrium configurations $C_1$, $C_2$ and $C_3$ scale with the dipole parameter $\chi$ and with the number of dipoles $N$. In Fig.\ref{fi:energies} we show the evolution of $E_{1,2,3}/\chi N$ as a function of the normalized distance $b$ for $r=3$. We observe an energy crossover between $E_2$ and $E_3$ at $b_c \approx 0.81$. Thus, for $b < b_c$ the energy $E_3$ is the minimal one, i.e. configuration $C_3$ is the ground state (GS) of the system, while for $b > b_c$ we have that $E_2$ is the minimal energy, so configuration $C_2$ is the GS. It is important noticing that, apart from $b$, the values of the energies $E_1$, $E_2$ and $E_3$ depend on the considered interaction order $r$. In this sense, we find that those values remain approximately constant for $r \ge 3$, such that the crossover value between $E_2$ and $E_3$ takes places at
$b \approx 0.81$ for $r \ge 3$.

In order to further clarify this point, we resort to a brute-force sampling method to find the GS.  The first step in our approach is to assume that the GS configuration of a ladder chain with a few dipoles (we use a ladder chain with $N=4$ dipoles) can be extended to systems with an arbitrary number of dipoles. In a second step, the potential $V$ of that system is evaluated in a huge number of random points $(u_1,..., u_N, v_1,..., v_N)$ uniformly distributed within the volume Vol$_0=(2 \pi)^{2N}$.
This sampling provides an estimate of the GS.
This sampling procedure is repeated for values of $b$ ranging in the interval $b\in [0.5, 1.5]$ and for different
interactions orders $r$.
In all cases, the results of this brute-force sampling method clearly predict the existence of two different GS configurations. On the one side, for
 $b\lesssim 0.81$ the GS is given by the configuration $C_3$, while for $b\gtrsim0.81$, the GS configuration is given by $C_2$.

In the next section we address the question about the energy crossover between the configurations $C_2$ and $C_3$ in the following way.  
Indeed, it is expected that this crossover indicates a change in the stability of those configurations. 
In order to study that stability change, we have to know the nature of the eigenvalues
$\{\lambda_n,  \forall n=1,..., 2N\}$ of the Hessian matrix
associated to the potential energy surface $V$ evaluated in those equilibria. However, those eigenvalues are the squared values of the 2N frequencies
$\{\omega_n=\sqrt{\lambda_n},  \forall n=1,..., 2N\}$
of the normal modes of the linearized dynamics associated to Hamiltonian \eqref{hamiEuler1}.
Therefore, in the next section we determine the dispersion relation associated to $C_2$ and $C_3$.
 As we already mentioned, the dispersion relation will be also used to address the question related to the order to which the interaction must be extended in order for the dynamics to be described correctly.

 \section{The linearized dynamics around the equilibrium $C_2$}
 \label{sec:linearizedC2}
Using the derivatives\ref{partialU}-\ref{partialV}, the Newtonian equations of motion associated to the Hamiltonian \ref{hamiEuler1} are
\begin{equation} 
\label{ecuMovi1}
\ddot u_n=-\frac{1}{I} \frac{\partial V}{\partial u_n},\quad \ddot v_n=-\frac{1}{I} \frac{\partial V}{\partial v_n}.
\end{equation} 
Let us considerer values of $b>b_c$, so that the GS is given by the equilibrium configuration $C_2=\{ u_n=0, v_n=\pi, \forall n \}$. 
In order to study the linear behavior of a certain nonlinear system around a stable equilibrium, it is convenient that such equilibrium be located at the origin. Therefore, we move
the equilibrium $C_2$ to the origin by applying the translation $\tau=\{v_n=v_n' + \pi, \forall n \}$ to the potential \ref{potential1}. Then, after this translation, and for small oscillations around that the translated $C_2$ that, we recall, is now located at $\{u_n=v'_n=0\}$,  the linear approximation to the equations of motion \ref{ecuMovi1} yields:
\begin{eqnarray}
\label{linearU}
\ddot u_n&=&-\w_0^2 \sum_{k=1}^r \frac{u_{n-k}+u_{n+k}+4 u_n}{k^3}-\w_0^2  \sum_{k=-r}^r \frac{(2b^2-k^2) v'_{n+k}+(b^2-2 k^2) u_n}{(b^2+k^2)^{5/2}},\\
\label{linearV}
\ddot v'_n&=&-\w_0^2 \sum_{k=1}^r \frac{v'_{n-k}+v'_{n+k}+4 v'_n}{k^3}-\w_0^2  \sum_{k=-r}^r \frac{(2 b^2 -k^2) u_{n+k}+(b^2-2 k^2) v'_n}{(b^2+k^2)^{5/2}},
\end{eqnarray}
where we have defined the frequency $\w_0=\sqrt{\chi/I}$. The solutions of Eqs. \eqref{linearU}-\eqref{linearV} are linear combinations of two different propagating plane waves in each sublattice \cite{solidsbook,A1021}. Therefore, we expect
solutions of the form
\begin{equation}
\label{waves}
u_n(t)=A(q) e^{i (q n a- \w t)}, \quad v'_n(t)=B(q) e^{i (q n a- \w t)},
\end{equation}
where $q$ is the wave number and $\w$ is the frequency.
Assuming pbc in both sublattices, the allowed values of the wave number $q$ are
\[
q=\frac{2 \pi m}{N a}, \quad m=1,..., N.
\]
The substitution of the ansatz \ref{waves} into Eqs.\ref{linearU}-\ref{linearV} yields
\begin{eqnarray}
\label{ecuU}
&& A(q) [\w^2-\w_0^2 \ F(q,r,b)] -B(q) \ \w_0^2 \ G(q,r,b)=0,\\
\label{ecuV}
&& A(q) \ \w_0^2 \ G(q,r,b)+B(q) [\w^2-\w_0^2 \ F(q,r,b)] =0,
\end{eqnarray}
where the functions $F(q,r,b)$ and $G(q,r,b)$ are
\begin{eqnarray}
\label{funf}
F(q,r,b)&=&\sum_{k=1}^r \frac{4+2\cos(q k a)}{k^3}+\frac{1}{b^3}+2\sum_{k=1}^r \frac{(b^2 -2 k^2)}{(b^2+k^2)^{5/2}},\\
\label{fung}
G(q,r,b)&=&\frac{2}{b^3}+2\sum_{k=1}^r \frac{(2 b^2 -k^2)}{(b^2+k^2)^{5/2} }  \cos(q k a).
\end{eqnarray}
In order to obtain nontrivial solutions for the Eqs.\ref{ecuU}-\ref{ecuV}, the determinant of the matrix of the $(A(q), B(q))$ coe\-ffi\-cients
has to vanish. Therefore, the relation between the frequency $\w$ and the wave number $q$, i.e., the dispersion relation, is
given through the following equation
\begin{equation}
\label{dispersion}
\w^4-2 \w^2 \w_0^2 F(q,r,b)+\w_0^4 (F(q,r,b)^2-G(q,r,b)^2)=0.
\end{equation}
The two solutions $\w_{\pm}$ of \ref{dispersion} read
\begin{equation}
\label{omega}
\w_{\pm}(q,b, r)=\w_0\sqrt{F(q,r,b) \pm G(q,r,b)},
\end{equation}
are the two bands of the dispersion relation. For a given distance $a$ and for all interaction order $r>1$, stable motion around $C_2$ implies that frequencies $\w_{\pm}(q,b, r)$ have to be positive for al $q$. This question will be address later.
When only the NN interaction is considered ($r=1$), and when the two chains are very far from each other ($b\rightarrow \infty$), we have that functions \ref{funf}-\ref{fung} become
\[
F(q,1,b\rightarrow \infty) \approx4+2\cos(q k a),\quad
G(q,1,b\rightarrow \infty)\approx 0,
\]

\noindent
therefore obtaining
\[
\w_-=\w_+\approx \w_0 \sqrt{4 + 2 \cos (q a)},
\]
\noindent
the expected single optical band of a linear chain of $N$ dipoles \cite{zampetaki,solidsbook}.

It is interesting to study in both dispersion bands the ratio between the amplitudes $A(q)$ and $B(q)$ of the waves propagating in each chain of the array. Inserting the expression \ref{omega} of the dispersion relation $\w_{\pm}$ in Eqs. \ref{ecuU}-\ref{ecuV} results in
\begin{equation}
\label{ratioAB}
\frac{A(q)}{B(q)} = \left\lbrace
\begin{array}{ll}
+1 \quad \textup{in the band} \quad \omega_+\\
-1 \quad \textup{in the band} \quad \omega_-.
\end{array}
\right.
\end{equation}
It is worth to note that this ratio does not depend on the distance $b$, nor on the interaction order $r$, nor on the wave number $q$. Therefore, in both bands $\w_{\pm}$ of the dispersion relation, the waves propagate in such a way that all dipoles of the array oscillate with the same amplitude. The linear behavior of the system in the neighborhood of the center ($q\approx 0$) and in the boundaries ($q\approx \pm \pi/a$) of the Brillouin zone can be found in the Appendix A.

\subsection{Evolution of the dispersion relation as a function of the interaction order $r$}
\begin{figure}[h]
\centerline{\includegraphics[scale=0.46]{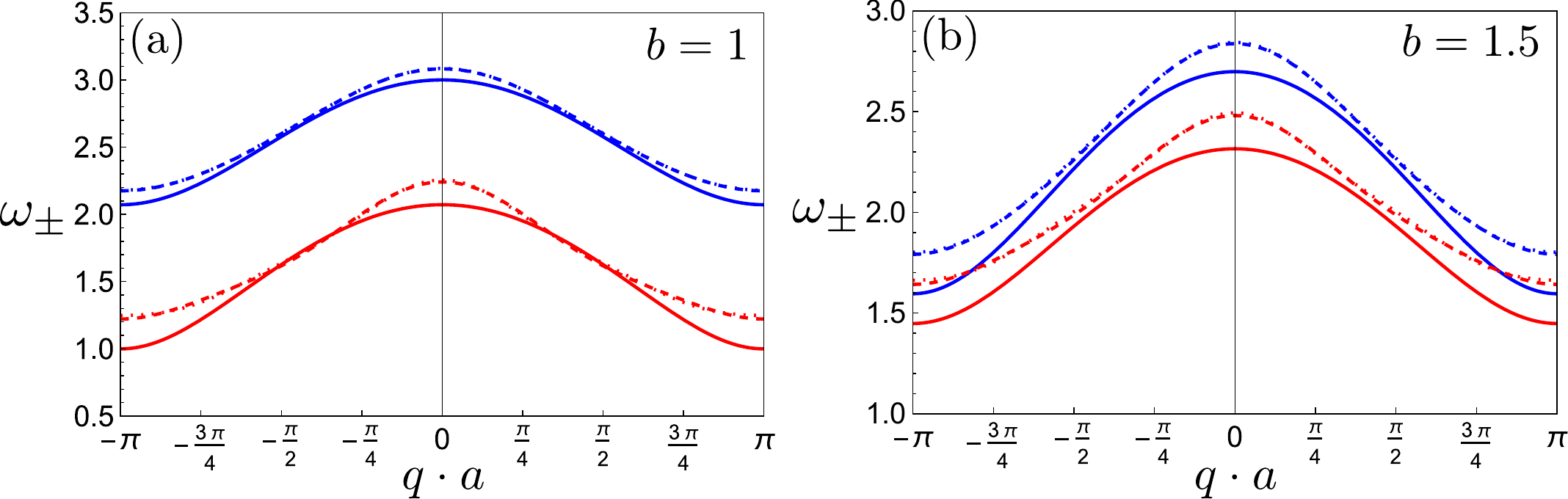}}
\caption{Dispersion relation $\w_{\pm}$ for $b=1$ and $b=1.5$. Solid lines stands for NN interaction $r=1$, while dashed and dotted lines
are for $r=3$ and $r=4$ interaction orders, respectively. Note that in both panels there is a complete overlap between the dashed and the dotted lines, indicating that, beyond $r=3$, the linear behavior of the system is not altered.}
\label{fi:dispersion}
\end{figure}

In addition to the distance parameter $b$ between the chains, the dispersion relation \ref{omega} depends on the considered interaction order $r$.
However, for a given value of $b$, it is expected that, for a certain value of the interaction order $r=r_c$, the shape of the two bands of the dispersion relation will not be affected significantly by the progressive inclusion of higher interaction orders $r>r_c$. In this way, for $b=1$ and $b=1.5$, we show in Fig.\ref{fi:dispersion} the dispersion relation \ref{omega} for $r=1, 3$ and 4. As we observe, when $r=3$ and $r=4$, the corresponding dispersion bands agree very well, which indicates that the inclusion of interaction orders beyond $r=3$ does not alter the linear behavior of the system.

\begin{figure}[h]
\centerline{\includegraphics[scale=0.6]{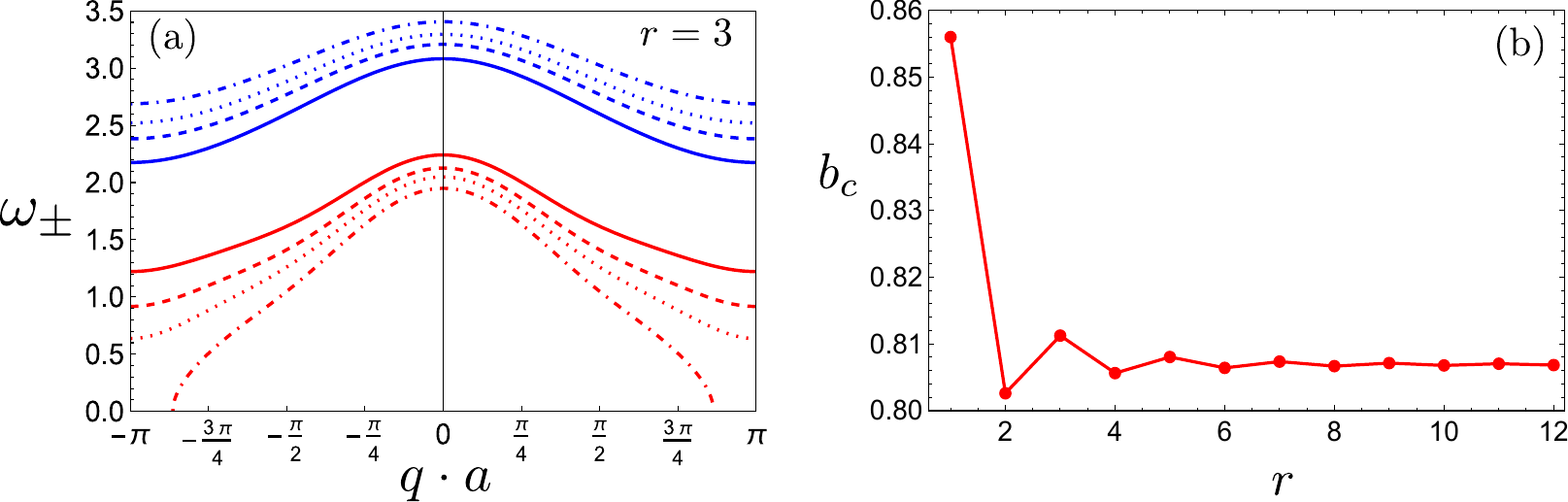}}
\caption{Left panel: Dispersion relation $\w_{\pm}$ for the interaction order $r=3$ and for $b=1$ (solid lines), $b=0.9$. (dashed lines), $b=0.85$ (dotted lines) and $b=0.8$ (dashed-dotted lines).  For $b>b_c=0.8112$ we have a complete positive linear spectrum, which indicates that, for $b>b_c$, the equilibrium $C_2$ is stable and it is the GS of the system.
Right panel: Evolution of the critical distance $b_c$ for different interaction orders $r$.}
\label{fi:dispersionEvolucion}
\end{figure}

As we observe in Fig.\ref{fi:dispersion}, the two bands of the linear spectrum are optic-like with the frequency $\w_{\pm}$ possessing a maximum for $q=0$ and a minimum for
$q = \pm \pi/a$. The numerical evaluation of the evolution of $\w_{\pm}$ with the distance $b$ indicates that the two bands separate for decreasing values of $b$. This fact is depicted in Fig.\ref{fi:dispersionEvolucion}(a) for the interaction order $r=3$. Therefore, and depending on the interaction order $r$, there appears for the lower band $\w_-$ a critical value $b_c$ such that, for $b<b_c$, part of the linear spectrum of $\w_-$ becomes complex. Because the minimum of $\w_-$ takes place at $q= \pm \pi/a$, the normal modes of the shortest wavelengths of $\w_-$ are the first to become complex. For $r=3$, we have a critical value $b_c \approx 0.8112$. Therefore, when $b<b_c$ the equilibrium configuration $C_2$ is no longer stable because part of the linear spectrum becomes complex and, therefore, $C_2$ cannot be the GS of the system. We notice that this is the expected behavior because Fig.\ref{fi:energies} shows a crossover between the energies of $C_2$ and $C_3$ for the same value $b=b_c$, as well as the brute force calculation of the GS indicates a change of the GS from the $C_2$ configuration to the $C_3$.

The value of the critical distance $b_c$ depends on the value of the interaction order $r$. For each value of $r$, we determine $b_c$ as the value of $b$ for which $F(q=\pi/a, r, b)=G(q=\pi/a, r, b)$.  Indeed, in Fig.\ref{fi:dispersionEvolucion}(b), where the evolution of $b_c$ as a function of $r$ is shown, we observe that $b_c$ asymptotically tends to the value $b_c\approx 0.8069$.

 \section{The linearized dynamics around the equilibrium $C_3$}
 \label{sec:linearizedC3}
In order to move the equilibrium $C_3$ to the origin, we apply the translation $\tau=\{u_n=u_n' + (-1)^n \pi/2, v_n=v_n' + (-1)^n \pi/2, \forall n \}$ to the potential \ref{potential1}. For small oscillations around the translated $C_3$ that is now located at $\{u_n'=v_n'=0, \forall n \}$, the linear approximation of the equations of motion \ref{ecuMovi1} yields:
\begin{eqnarray}
\label{linearU2}
\ddot u'_n&=&2 \w_0^2 \sum_{k=1}^r (-1)^k \frac{u'_{n-k}+u'_{n+k}+u'_n}{k^3}-\w_0^2 \sum_{k=-r}^r (-1)^k \frac{(2b^2-k^2) u'_n+(b^2-2 k^2) v'_{n+k}}{(b^2+k^2)^{5/2}},\\
\label{linearV2}
\ddot v'_n&=&2\w_0^2 \sum_{k=1}^r (-1)^k \frac{v'_{n-k}+v'_{n+k}+v'_n}{k^3}-\w_0^2  \sum_{k=-r}^r (-1)^k \frac{(2 b^2 -k^2) v'_n+(b^2-2 k^2) u'_{n+k}}{(b^2+k^2)^{5/2}}.
\end{eqnarray}
After the substitution of the ansatz \ref{waves} in Eq.\ref{linearU2}-\ref{linearV2}, we
obtain the following equations
\begin{eqnarray}
\label{ecuU2}
&& A(q) [\w^2-\w_0^2 \ M(q,r,b)] -B(q) \ \w_0^2 \ N(q,r,b)=0,\\
\label{ecuV2}
&& A(q) \ \w_0^2 \ M(q,r,b)+B(q) [\w^2-\w_0^2 \ N(q,r,b)] =0,
\end{eqnarray}
where the functions $M(q,r,b)$ and $N(q,r,b)$ are
\begin{eqnarray}
\label{funm}
M(q,r,b)&=&-2\sum_{k=1}^r (-1)^k \frac{1+2\cos(q k a)}{k^3}+
\frac{2}{b^3}+2\sum_{k=1}^r (-1)^k \frac{(2b^2-k^2)}{(b^2+k^2)^{5/2}},\\
\label{funn}
N(q,r,b)&=&\frac{1}{b^3}+2\sum_{k=1}^r (-1)^k \frac{(b^2 -2k^2)}{(b^2+k^2)^{5/2} }  \cos(q k a).
\end{eqnarray}
From the determinant of the coefficient matrix of Eqs.\ref{ecuU2}-\ref{ecuV2} equated to zero, we have that the dispersion relation is given by the following equation
\begin{equation}
\label{dispersion2}
\w^4-2 \w^2 \w_0^2 M(q,r,b)+\w_0^4 (M(q,r,b)^2-N(q,r,b)^2)=0,
\end{equation}
such that the two bands of the dispersion relation are the solutions $\w_{\pm}$ of \ref{dispersion2},
\begin{equation}
\label{omega2}
\w_{\pm}(q,b, r)=\w_0\sqrt{M(q,r,b) \pm N(q,r,b)}.
\end{equation}

\noindent
The substitution of this expression \ref{omega2} in Eqs. \ref{ecuU2}-\ref{ecuV2} give us the same ratio between the amplitudes $A(q)$ and $B(q)$ of the plane waves \ref{waves} as the ratio \ref{ratioAB} we have found for the $C_2$ configuration. Therefore, the linear behavior of the array in this configuration is very similar to the one observed in the $C_2$ configuration. That is, in both bands $\w_{\pm}$, all dipoles of the array oscillate with the same amplitude.  The linear behavior of the system in the neighborhood of the center ($q\approx 0$) and in the boundaries ($q\approx \pm \pi/a$) of the Brillouin zone can be found in the Appendix B.

\begin{figure}[h]
\centerline{\includegraphics[scale=0.47]{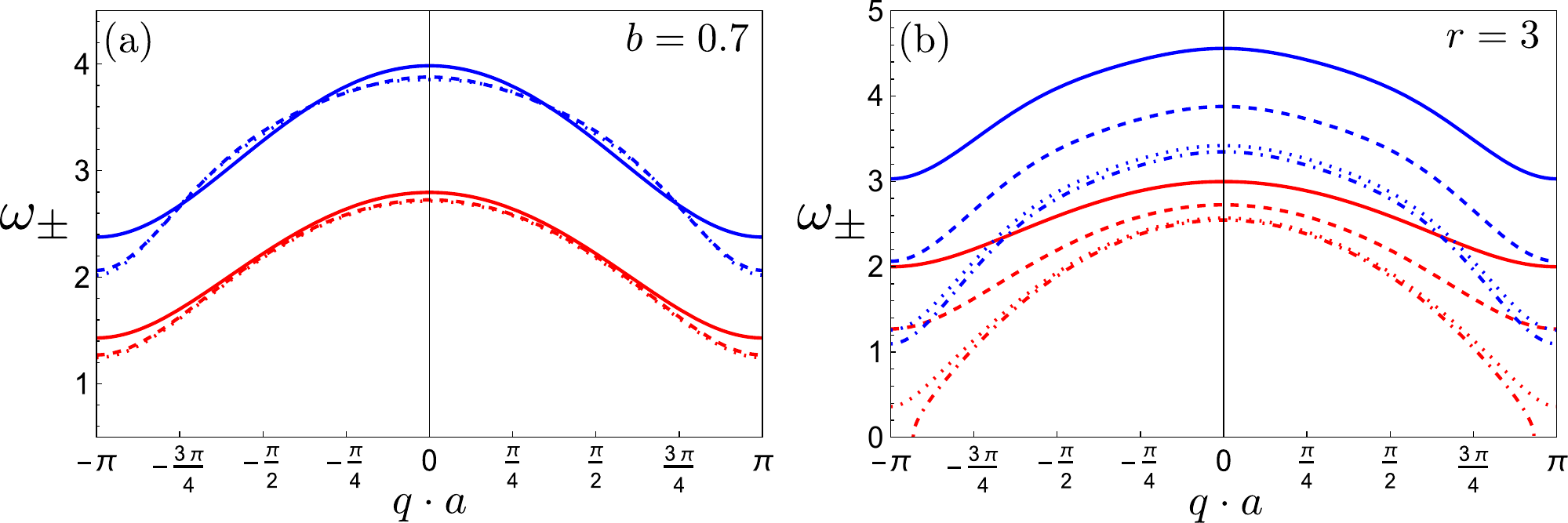}}
\caption{Left panel: Dispersion relation $\w_{\pm}$ given by Eq.\ref{dispersion2} for $b=0.7$. Solid lines stand for NN interaction $r=1$, while dashed and dotted lines
are for $r=3$ and $r=4$ interaction orders, respectively. Note that there is a very good agreement between the $r=3$ and $r=4$ cases, indicating that, beyond $r=3$, the linear dynamics is not altered significantly. Right panel: Dispersion relation $\w_{\pm}$ for the interaction order $r=3$ and for $b=0.6$ (solid lines), $b=0.7$. (dashed lines), $b=0.8$ (dotted lines) and $b=0.82$ (dashed-dotted lines).  For $b<b_c=0.8112$ we have a complete positive linear spectrum, which indicates that, for $b<b_c$, the equilibrium $C_3$ is stable and it is the GS of the system.}
\label{fi:dispersion2}
\end{figure}

The dispersion relation \ref{omega2} depends on the considered interaction order $r$. As we observe in Fig.\ref{fi:dispersion2}(a) for $b=0.7$ and for $r=1, 3$ and 4, the  dispersion bands \ref{omega2} for $r=3$ and $r=4$ agree very well, which indicates that, as in the previous case around $C_2$, the inclusion of interaction orders beyond $r=3$ has only a minor influence on the linearized dynamics.
We observe in Fig.\ref{fi:dispersion2}(a) for $b=0.7$ that the two bands of \ref{omega2} are optic-like with a maximum and minimum values at $q=0$ and $q=\pm \pi/a$, respectively. As we already mentioned, for each interaction order, we expect the existence of a critical $b_c$ such that for $b>b_c$ the equilibrium $C_3$ will not be the GS.
Fig. \ref{fi:dispersion2}(b) shows the evolution of the dispersion bands $\omega_\pm$ with the distance $b$ for the interaction order $r=3$. We find that, for the critical value $b_c \approx 0.8112$, the linear spectrum of the lower band $\omega_-$ begins to be complex. Thus, for $b>b_c$ the equilibrium configuration $C_3$ is no longer stable, and the $C_2$ configuration becomes the GS of the system. 

Following the same procedure as for the equilibrium $C_2$, we find numerically, for each interaction order $r$, the value $b_c$ as the value of $b$ for which $M(q=\pi/a, r, b)=N(q=\pi/a, r, b)$. As expected, we obtain the same results as in Fig.\ref{fi:dispersionEvolucion}(b). Therefore,  at any order $r$, for $b<b_c \ (b>b_c))$  the linear spectrum around $C_3 \ (C_2)$ is positive and the equilibrium $C_3 \ (C_2)$ represents the GS of the system.

\section{ENERGY TRANSFER UNDER SINGLE SITE EXCITATION}
In this section we study the time propagation of single site excitations. More precisely, starting from the GS configuration, we excite one dipole supplying it with an excess of kinetic energy $\Delta K$.
In the previous sections we have shown that extending the interactions to orders $r>3$
does not significantly alter the linearized dynamics. Therefore, we assume that the complete dynamics will not be dramatically affected either by the inclusion of terms of order $r>3$. 
Thus, in all our calculations we extend the dipole interaction up to the order $r=3$.
In this way, each dipole will interact with its thirteen nearest neighbors.
We can visualized those thirteen interactions in Fig.\ref{fi:chain1} by considering that dipole $u_i$ (or dipole $v_i$) interacts with the remaining thirteen dipoles of Fig.\ref{fi:chain1}.

For $r=3$ the critical distance between the chains is $b_c \approx 0.8112$. Therefore, being the energy of the GS denoted as $E_{GS}$, when $b > b_c \approx 0.8112$ we have that $E_{GS}=E_2$ (see Eq.\ref{enerC2}), while when $b < b_c \approx 0.8112$ we have that $E_{GS}=E_3$ (see Eq.\ref{enerC3}). Without loss of generality, we excite the first dipole of the lower chain. Thus, at $t=0$, the initial conditions of the array are
\begin{eqnarray}
\nonumber
v_n(0)&=&\pi, \quad P_n(0)=0, \quad \forall \ n,\\
\label{ini}
u_n(0)&=&Q_n(0)=0, \quad \forall \ n \ne 1,\\
\nonumber
u_1(0)&=&0, \quad Q_1(0)=\sqrt{2 \Delta K}.
\end{eqnarray}
In this study we use $N=100$ dipoles in each chain. To carry out this study, it is convenient to use 
a dimensionless version of the Hamiltonian \ref{hamiEuler1}. Then, scaling the energy $E$ in terms of the dipole
parameter $\chi$ as $E'=E/\chi$,  we readily arrive at the following dimensionless Hamiltonian
\begin{equation}
\label{hamiEuler2}
E'=\frac{{\cal H}}{\chi}=\sum_{n=1}^{N} \frac{Q_n'^2+P_n'^2}{2} + V(u_1,..., u_N, v_1,..., v_N), 
\end{equation}
where $Q_n'=d u_n/d t'$, $P_n'=d v_n/d t'$ and $t'=t/\tau$ is a dimensionless time measured in units of $\tau=\sqrt{I/\chi}$.
In order to simplify the notation, we omit the primes in the Hamiltonian \ref{hamiEuler2}.
With the initial conditions \ref{ini}, the Hamiltonian equations of motion for different values of $b$ and $\Delta K$ are integrated from an initial time $t=0$, up to a final time $t_f$ by means of the SABA$_2$ symplectic integrator \cite{A1095,A908}.

At this point, and using Eqs.\ref{v1}-\ref{v2}-\ref{v3}, we define the local energy stored at a given time $t$ in each dipole as
\begin{eqnarray}
\label{localu}
E_n^u(t)&=&\frac{Q_n(t)^2}{2}+\frac{1}{2}\sum_{\substack{j=-3\\j\ne 0}}^3V_1(u_n(t),u_{n+j}(t))+\frac{1}{2}\sum_{j=-3}^3V_3(u_n(t),v_{n+j}(t))-\frac{E_{\rm GS}}{2 N},\\[2ex]
\label{localv}
E_n^v(t)&=&\frac{P_n(t)^2}{2}+\frac{1}{2}\sum_{\substack{j=-3\\j\ne 0}}^3V_2(v_n(t),v_{n+j}(t))+\frac{1}{2}\sum_{j=-3}^3V_3(u_{n+j}(t),v_n(t))-\frac{E_{\rm GS}}{2 N}.
\end{eqnarray}
With this definition of the local energies, the GS is shifted to zero, such that the total energy of the system is $\Delta K$. Therefore, the total energy located at time $t$ in each chain of the array is computed as
\begin{equation}
\label{totaluv}
k_u(t)=\sum_{n=1}^N E_n^u(t), \qquad k_v(t)=\sum_{n=1}^N E_n^v(t),\qquad k_u(t)+k_v(t)=\Delta K.
\end{equation}
By using the expressions \ref{totaluv} we define the participation functions $\Pi_u(t)$ and $\Pi_v(t)$ as
\begin{eqnarray}
\label{piu}
\Pi_u(t)&=&\frac{1}{N-1}\left(\frac{k_u^2(t)}{\sum_{n=1}^N E_n^u(t)^2}-1\right),\\[2ex]
\label{piv}
\Pi_v(t)&=&\frac{1}{N-1}\left(\frac{k_v^2(t)}{\sum_{n=1}^N E_n^v(t)^2}-1\right).
\end{eqnarray}
When at a given time $t$ the total energy stored in one of the chains is maximally localized (carried by a single dipole), the value of the corresponding participation function is zero. On the contrary, when in any of the chains there is a complete equipartition of the energy, (i.e., $E_n^u(t)=k_u(t)/N$ or $E_n^v(t)=k_v(t)/N, \ \forall n$), the corresponding participation function is one.

In our system, a global description based on the parameters $\Delta K$ and $b$ of a dynamical process such as the energy transfer can be obtained using average values of the functions \ref{totaluv}-\ref{piu}-\ref{piv} calculated in convenient time windows. Then, to follow the propagation and the distribution of the initially localized excitation $\Delta K$ between the two chains, we define the normalized average energies  $\langle K_u\rangle$ and $\langle K_v\rangle$ as
\begin{eqnarray}
\label{prome}
\langle K_u\rangle&=&\frac{\langle k_u\rangle}{\langle k_u\rangle+\langle k_v\rangle}, \qquad \langle K_v\rangle=\frac{\langle k_v\rangle}{\langle k_u\rangle+\langle k_v\rangle},
\end{eqnarray}
\noindent
where $\langle k_u\rangle$ and $\langle k_v\rangle$ are the average values of $k_u(t)$ and $k_v(t)$ in the time interval $\Delta t:=[t_i, t_f]$ calculated as,
\begin{eqnarray}
\langle k_u\rangle&=&\frac{1}{t_f-t_i} \int_{t_i}^{t_f}k_u(t) dt,
\qquad \langle k_v\rangle=\frac{1}{t_f-t_i} \int_{t_i}^{t_f}k_v(t) dt.
\end{eqnarray}
Similarly, in order to quantify the localization of the energy along each chain, we define the time averaged participation functions $\langle\Pi_u \rangle$ and $\langle\Pi_v \rangle$ as
\begin{equation}
\label{prome2}
\langle \Pi_u\rangle=\frac{1}{t_f-t_i} \int_{t_i}^{t_f}\Pi_u(t) dt,\qquad
\langle \Pi_v\rangle=\frac{1}{t_f-t_i} \int_{t_i}^{t_f}\Pi_v(t) dt.
\end{equation}

\medskip
The choice of the value of the final integration time $t_f$ is a delicate issue. On the one hand, it is expected that, for sufficiently larges values of $t_f$, both chains will reach thermal equilibrium, which would be characterized by $\langle K_u\rangle=\langle K_v\rangle=\langle K_{\rm eq}\rangle \approx 0.5$ and by $\langle \Pi_u\rangle=\langle \Pi_v\rangle=\langle \Pi_{\rm eq}\rangle$, with $\langle \Pi_{\rm eq}\rangle$ a certain stationary value.
On the other hand, and depending on the particular values of $\Delta K$ and $b$, the system will go through (possibly) dynamical phases with different characteristic time scales before thermal equilibrium is reached.
To uncover those dynamical scenarios, we will compute $\langle K_{u,v}\rangle$ and $\langle \Pi_{u,v}\rangle$ for several values of $\Delta K$ and $b$, and for different time windows $\Delta t:=[t_i, \ t_f]$.

\subsection{Energy transfer in the $C_2$ configuration}
We start by assuming that the $C_2$ configuration is the GS of the system (i.e., $b>b_c \approx 0.8112$, $E_{\rm GS}=E_2$, Eq.\ref{enerC2}). Investigating the parameter plane $(\Delta K, b)$ in the excess energy interval $1 \le \Delta K \le 15$ (in units in of $\chi$) and in the distance interval $0.9 \le b \le 2.5$, we generate two-dimensional color maps with the values of $\langle K_u\rangle$, $\langle \Pi_u\rangle$ and $\langle \Pi_v\rangle$ determined for different time-windows $\Delta t$. More specifically, we choose the time intervals $\Delta t_1:=[0, \ 10^5]$, $\Delta t_2:=[4\times10^5, \ 5\times 10^5]$ and $\Delta t_3:=[9\times10^5,\ 10^6]$, which allows us to study the energy transfer process from early to late times. 

\begin{figure}[h]
\centerline{\includegraphics[scale=0.35]{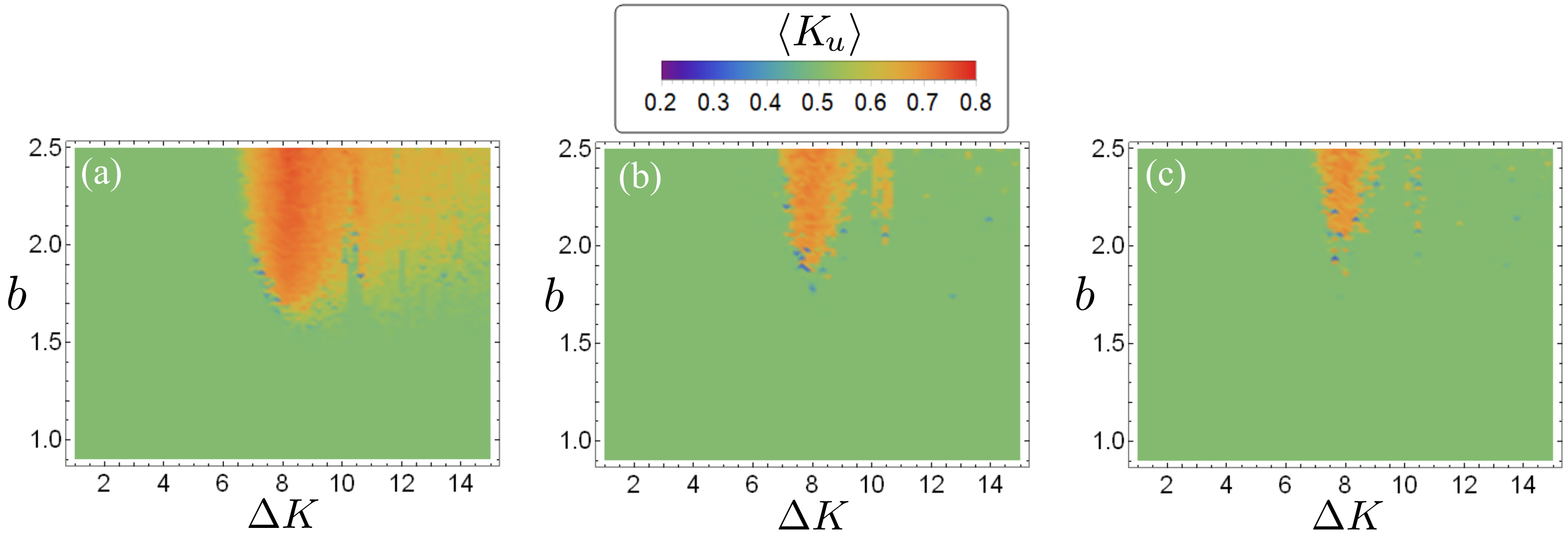}}
\caption{Colour maps of the time averaged normalized energy $\langle K_u\rangle$ stored in the active $u-$chain for three different time intervals: (a) $\Delta t_1:=[0, \ 10^5]$, (b)  $\Delta t_2:=[4\times10^5, \ 5\times 10^5]$ and (c) $\Delta t_3:=[9\times10^5,\ 10^6]$.}
\label{fi:colorMaps31}
\end{figure}

The color map depicted in Fig.\ref{fi:colorMaps31}(a) shows the values of $\langle K_u\rangle$ calculated during the time interval $\Delta t_1:=[0, 10^5]$, that is, in the earliest stage of the time evolution of the system.
We observe a dominant green coloured region for which the energy of the system is evenly distributed between the two chains, $\langle K_u\rangle=\langle K_v\rangle\approx 0.5$. The remaining region of the map with values $\Delta K \gtrsim 7$ and $b \gtrsim 1.5$ is dominated by orange-red colors indicating that, for those values of the parameters, the system energy is largely stored in the active $u-$chain.
It is important to notice that the borders between these two regions are not smooth. Therefore, it is not possible to predict the final outcome of a given initial condition belonging to those borders, i.e., whether or not the system will end up in energy equipartition.

In the intermediate and late time stages $\Delta t_2:=[4\times10^5, \ 5\times 10^5]$ and $\Delta t_3:=[9\times10^5,\ 10^6]$ (see Figs.\ref{fi:colorMaps31}(b)-(c)), there is a progressive increase in size of the green region in the parameter plane $(\Delta K, b)$ where the system energy is evenly distributed between both chains. It is worth to highlight the fact that, as Fig. \ref{fi:colorMaps31}(c) shows, even up to the maximum computation time $t_f=10^6$ used in this study, there still persists a region between $b\gtrsim 2$ and $7 \lesssim \Delta K \lesssim 9$ where the energy equipartition regime between the two chains of the array has not been reached.

\begin{figure}[h]
\centerline{\includegraphics[scale=0.35]{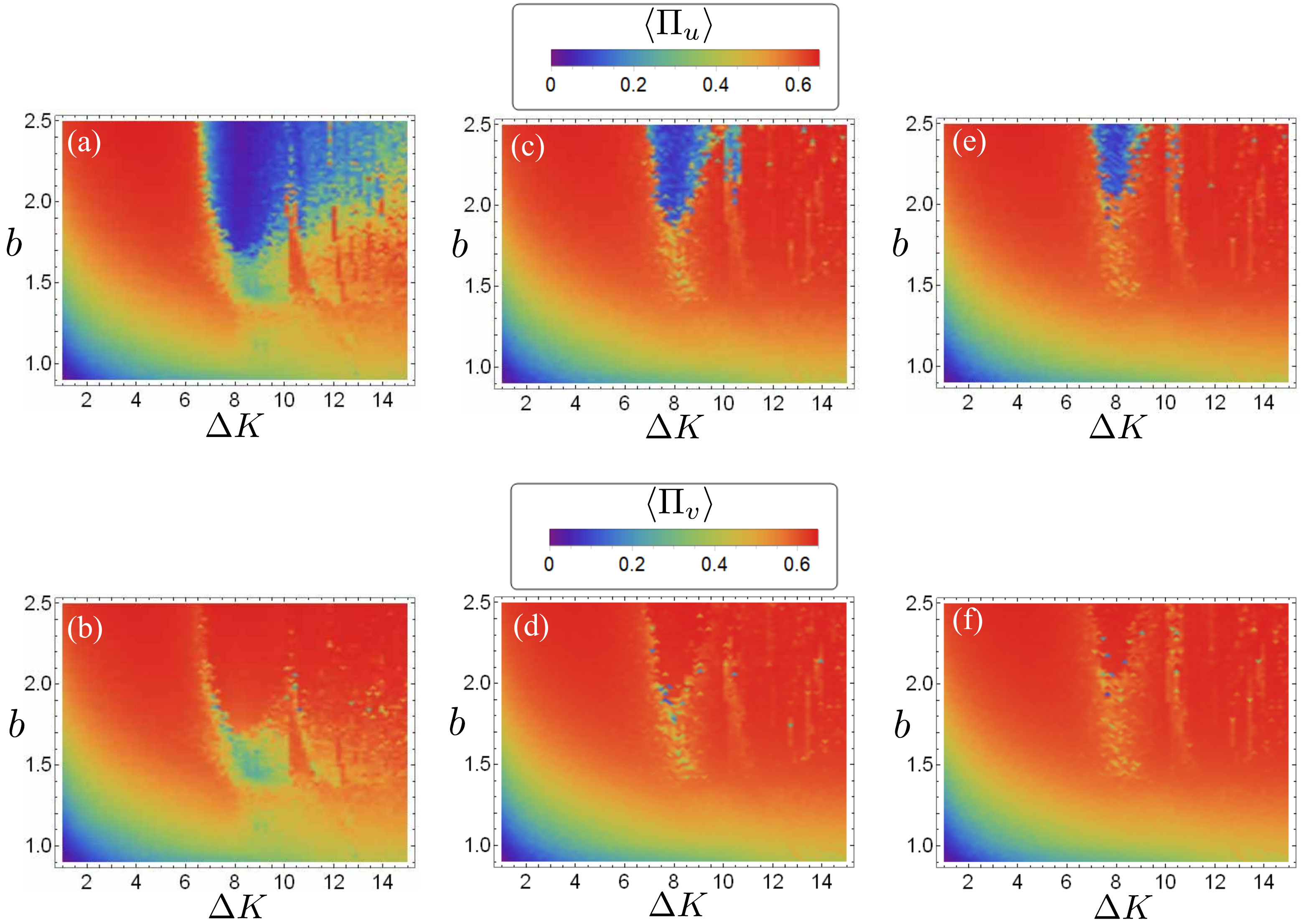}}
\caption{Colour maps of the time averaged participation functions $\langle \Pi_u\rangle$ (upper row) and $\langle \Pi_v\rangle$ (lower row). In panels (a)-(b) the quantities $\langle \Pi_u\rangle$ and $\langle \Pi_v\rangle$ are determined for the time interval $\Delta t_1:=[0, \ 10^5]$. In panels (c)-(d) and in panels (e)-(f) the same quantities are determined for the time intervals $\Delta t_2:=[4\times10^5, \ 5\times 10^5]$ and $\Delta t_3:=[9\times10^5,\ 10^6]$, respectively.}
\label{fi:colorMaps32}
\end{figure}

\begin{figure}[h]
\centerline{\includegraphics[scale=0.5]{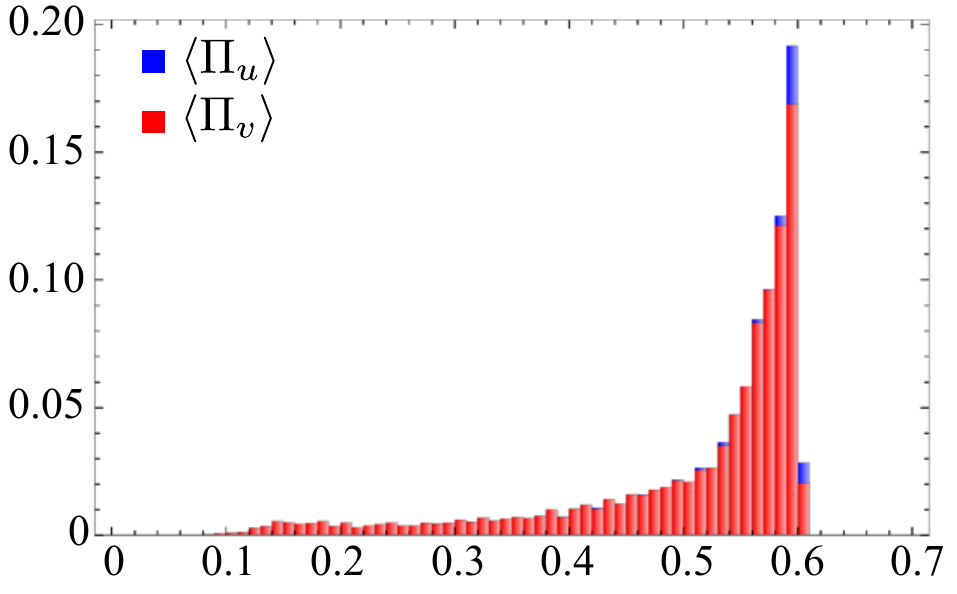}}
\caption{Probability distribution function (PDF) of the values
of $\langle \Pi_{u,v}\rangle$ for the latest time interval $\Delta t_3:=[9\times10^5,\ 10^6]$ of the color maps of Figs.\ref{fi:colorMaps32}(e)-(f).}
\label{fi:histogram}
\end{figure}

Regarding the way the energy is distributed in each chain, we illustrate in Fig.\ref{fi:colorMaps32} the time evolution of the time averaged participation functions for both the active $u-$chain $\langle \Pi_u\rangle$ (Fig.\ref{fi:colorMaps32} upper row) and the passive $v-$chain $\langle \Pi_v\rangle$ (Fig.\ref{fi:colorMaps32} lower row) calculated for the same time intervals $\Delta t_1:=[0, \ 10^5]$, $\Delta t_2:=[4\times10^5, \ 5\times 10^5]$ and $\Delta t_3:=[9\times10^5,\ 10^6]$. 
The color scale in Fig.\ref{fi:colorMaps32} indicates that, in all cases, the energy stored within each chain is far from the perfect equipartition regime given by $\langle \Pi_u\rangle=\langle \Pi_v\rangle=1$. Indeed, Fig.\ref{fi:histogram} displays the probability distribution function (PDF) of the values of $\langle \Pi_u\rangle$ and $\langle \Pi_v\rangle$ calculated for $\Delta t_3:=[9\times10^5,\ 10^6]$ and appearing in the maps of Figs.\ref{fi:colorMaps32}(e)-(f). We notice that the maximum values attained by $\langle \Pi_{u,v}\rangle$ in the parameter plane $(\Delta K, b)$ are roughly around 0.6, which corresponds to the red color in Fig.\ref{fi:colorMaps32}.  Moreover, this value is very close to the thermal equilibrium value of the participation function of a single linear chain obtained in \cite{chain} using Boltzmann statistics. 
Therefore, within the red color regions in Fig.\ref{fi:colorMaps32}, we may say that both chains and, therefore, the total system, have reached the thermal equilibrium, which is characterized by a value $\langle \Pi_{u,v}\rangle=\langle \Pi_{eq}\rangle\approx 0.6$.
As expected, during the intermediate $\Delta t_2$ and long times $\Delta t_3$ stages, there is a significant and progressive increase in size of the red regions in the plane $(\Delta K, b)$ where both chains are in thermal equilibrium.


\noindent
Besides the thermal equilibrium regions, we can distinguish in the color panels of Fig.\ref{fi:colorMaps32} several regions where the blue color dominates (low values of 
$\langle \Pi_u\rangle$ and $\langle \Pi_v\rangle$), which indicates a strong energy localization within the corresponding chain.
Indeed, in the active $u-$chain there is a region located in the upper part of the
maps of Figs.\ref{fi:colorMaps32}(a)-(c)-(e) for $b \gtrsim 1.5$ where the blue color dominates. We denote this region as $U$-zone.
It is worth to note that the size and the evolution of the $U$-zone during the different time windows match very well with the orange-red regions appearing in the color maps of $\langle K_u\rangle$ of Fig.\ref{fi:colorMaps31}.
Therefore, since for the parameter values of the $U$-regions there is no energy equipartition between the two chains (the energy stored in the active $u-$chain persists strongly localized), the system, up to the maximum time  $t_f=10^6$ considered in this study, is not able to reach the thermal equilibrium.
\begin{figure}[h]
\centerline{\includegraphics[scale=0.8]{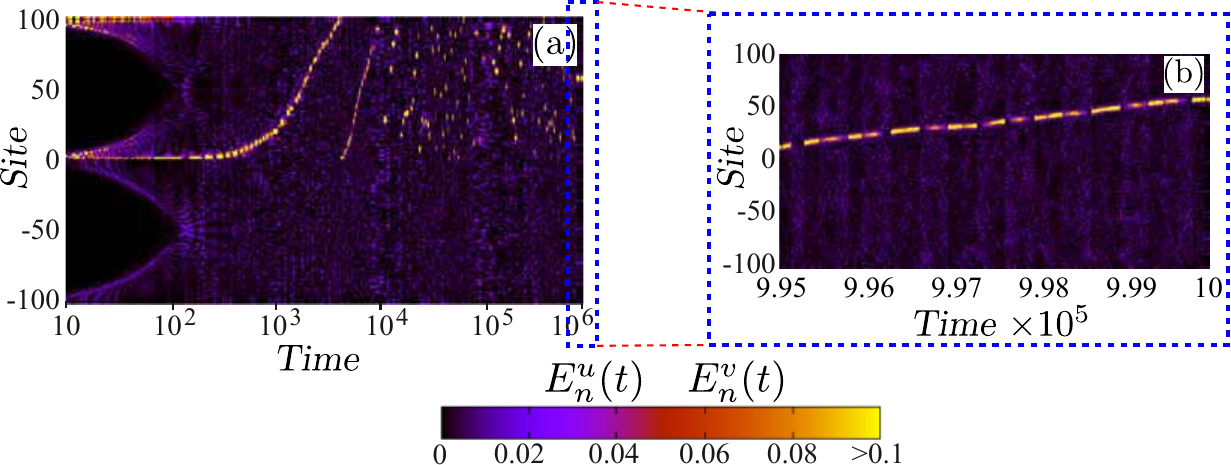}}
\caption{Time evolution of the normalized local kinetic energies $E_n^u(t)$ and $E_n^v(t)$ of the dipoles for $(\Delta K, b)=(8, 2.2)$. The positive values of the vertical axis stand for the $N=100$ dipoles of the $u$-chain, while the negative ones stand for the $N=100$ dipoles of the $v$-chain. (a) Evolution for the complete propagation time. (b) Magnification of the 5000 final time units. Note that in panel (a) a logarithmic scale is used for the horizontal axis.}
\label{fi:localEnergyEvolution2}
\end{figure}

An explanation of the behavior of the system in the $U$-zone can be obtained from the ana\-ly\-sis of the behavior of a single trajectory with appropriate initial conditions in that region. More precisely, we visualize how the excess energy $\Delta K$  is distributed along each of the dipoles by computing a two dimensional color map with the time evolution of the normalized local energies $\{(E_n^u(t)/\Delta K, E_n^v(t)/\Delta K), n=1,..., N\}$ (see Eqs.\ref{localu}-\ref{localv}). In particular, for a trajectory
with parameter values $(\Delta K, b)=(8, 2.2)$, the color map of Fig. \ref{fi:localEnergyEvolution2}(a) shows the presence of a chaotic breather that propagates
in the active $u$-chain and where a significant amount of the total energy $\Delta K$ is localized. As we can observe in the magnification of Fig.\ref{fi:localEnergyEvolution2}(b), where the 5000 final time units of Fig.\ref{fi:localEnergyEvolution2}(b) are shown, the breather clearly persists up to the the maximal computing time $t=10^6$.
Therefore, this chaotic breather prevents both the energy equipartition between the two chains and the thermal equilibrium in the $u$-chain.
In other words, the presence of chaotic breathers explains the existence of the orange-red region in the color maps of Fig. \ref{fi:colorMaps31} and the $U$-zone in the color maps of Figs. \ref{fi:colorMaps32}(a), (c) and (e).
It is expected that, in order to observe the energy equipartition between both chains and a global thermal equilibrium, we should go to times long enough for the breather to fade away. 

An additional region where the blue color dominates can be observed in the lower left corner of the maps of Figs.\ref{fi:colorMaps32} for low values of $\Delta K$ and $b$.
The size of this region, named as $L$-zone, remains roughly constant during the considered time windows.
Note that, despite the two chains are in mutual energy equipartition (see the color maps of Fig.\ref{fi:colorMaps31}), neither of the two chains reaches the thermal equilibrium.

\begin{figure}[h]
\centerline{\includegraphics[scale=0.35]{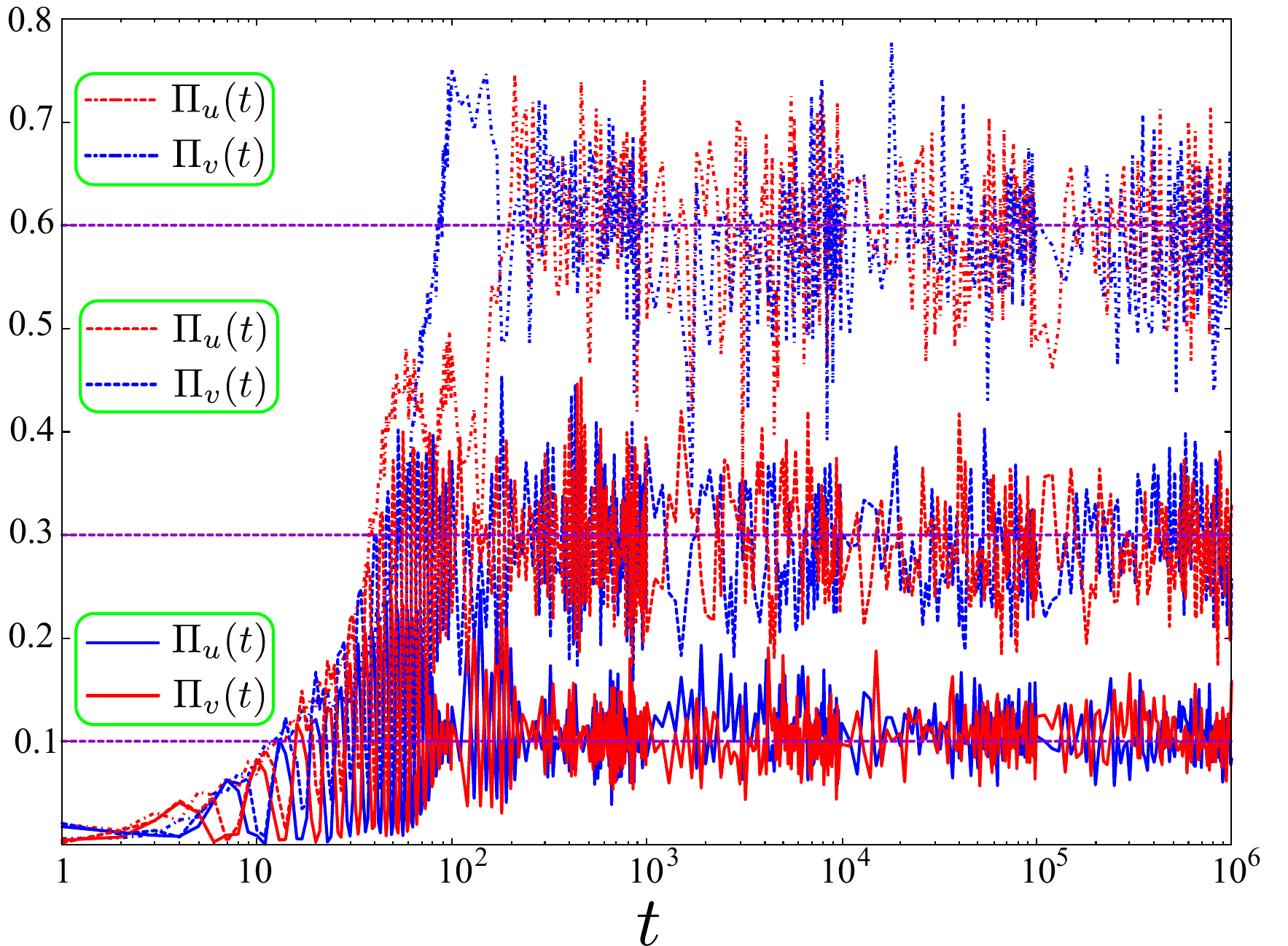}}
\caption{Time evolution of the participation functions $\Pi_{u,v}(t)$ of three different trajectories with $(\Delta K, b)=(1.5, 0.9)$ (solid lines), $(\Delta K, b)=(4, 1)$ (dashed lines) and $(\Delta K, b)=(1.5, 2)$ (dashed-dotted lines).}
\label{fi:participationEvolution}
\end{figure}

To explain the behavior of the energy transfer in the $L$-zone, we turn again to study the time evolution
of particular trajectories. Indeed, the participation functions $\Pi_{u,v}(t)$ (see Eqs.\ref{piu}-\ref{piv}) for a orbit with $\Delta K=1.5$ and $b=0.9$ depicted in Fig.\ref{fi:participationEvolution} show that, up to $t_f=10^6$,  the functions $\Pi_{u,v}(t)$ fluctuate around a constant
nonzero value $\langle \Pi_{u,v}\rangle \approx 0.1$, which is well below the thermal equilibrium value $\langle \Pi_{eq}\rangle \approx 0.6$.
In contrast to this behavior, for the same excess energy $\Delta K=1.5$ and a larger distance $b=2$ between the chains (this values are within a red region in  Fig.\ref{fi:colorMaps32}), we observe in Fig.\ref{fi:participationEvolution} that $\Pi_{u,v}(t)$ fluctuate around the thermal equilibrium value $\langle \Pi_{eq}\rangle \approx 0.6$. For $\Delta K=4$ and $b=1$, we see in Fig.\ref{fi:participationEvolution} that $\Pi_{u,v}(t)$ fluctuate around a intermediate value $\langle \Pi_{u,v}\rangle \approx 0.3$.

\begin{figure}[h]
\centerline{\includegraphics[scale=0.28]{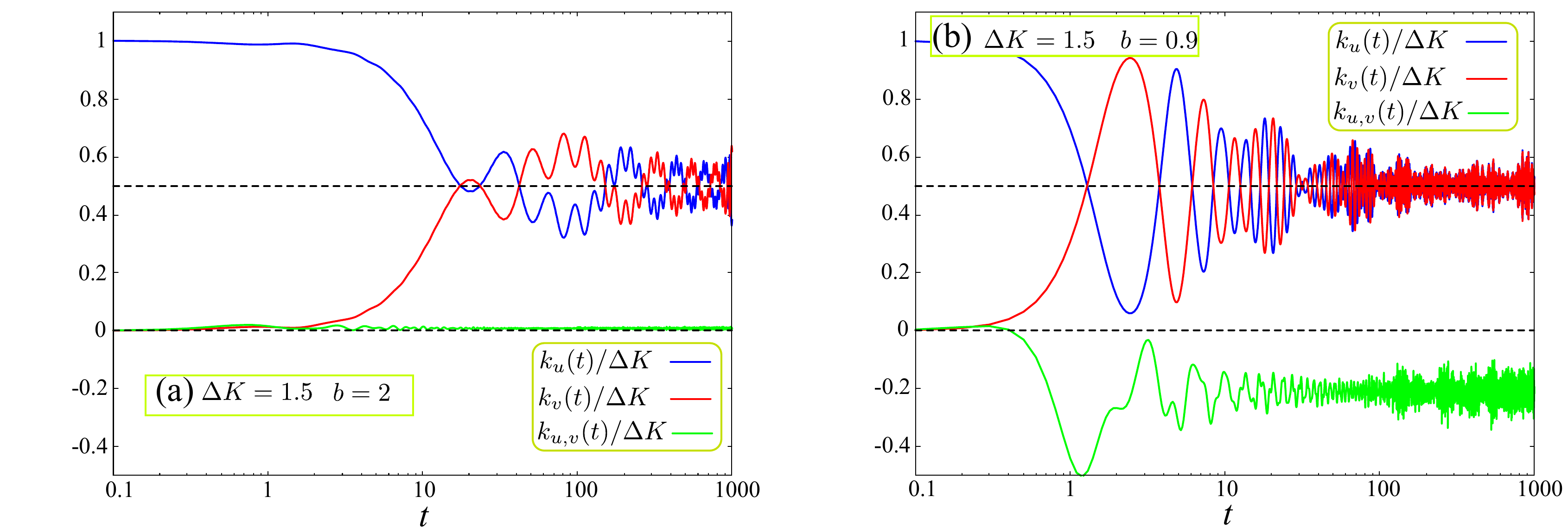}}
\caption{Time evolution of the normalized energy $k_u(t)/\Delta K$ stored in the $u$-chain (blue line), the  normalized energy  $k_v(t)/\Delta K$ stored in the $v$-chain (red line) and the normalized mutual interaction energy $k_{u,v}(t)/\Delta K$ between the chains (green line). These quantities have been computed for (a) $(\Delta K, b)=(1.5, 2)$ and (b) $(\Delta K, b)=(1.5, 0.9)$.}
\label{fi:trayectories}
\end{figure}

The remarkable robustness of the lower left blue area of the map of Fig.\ref{fi:colorMaps32}, suggets that, up to the maximum computation time $t_f=10^6$ used here, the thermalization of the system is delayed.
At this point, it is useful to study the time evolution of the total energies $k_u(t)$ and $k_v(t)$ stored in the $u$-chain and in the $v$-chain (see Eqs.\ref{localu}-\ref{localv}), respectively, as well as the mutual interaction energy $k_{u,v}(t)$ between the chains. According to Eqs.\ref{localu}-\ref{totaluv}, that mutual energy $k_{u,v}(t)$ is given by
\begin{equation}
\label{mutual}
k_{u,v}(t)=\sum_{n=1}^N\sum_{j=-3}^3V_3(u_n(t),v_{n+j}(t)).
\end{equation}
For $(\Delta K, b)=(1.5, 0.9)$ and $(\Delta K, b)=(1.5, 2)$, in Fig.\ref{fi:trayectories} the time evolution of the normalized quantities $k_u(t)/\Delta K$, $k_v(t)/\Delta K$ and $k_{u,v}/\Delta K$ of the corresponding trajectories is shown. In both examples, it is worth noticing that, shortly after the excitation, the system reaches, in average, the energy equipartition between the two chains.
However, the energy transfer mechanism in each sample trajectory is different.
Indeed, when the distance $b$ between the chains is large ($b=2$, Fig.\ref{fi:trayectories}(a)), the mutual energy $k_{u,v}(t)$ is small and positive, i.e., the two chains weakly repel each other.
Therefore, the interaction taking place between dipoles belonging to the same chain plays the most important role in the dynamics. Then, since the total energy $\Delta K$ at stake is small and the chains are in mutual energy equipartition, the thermalization of the system is rather fast.

When the distance between the chains is small ($b=0.9$, Fig.\ref{fi:trayectories}(b)), the mutual interaction energy $k_{u,v}(t)$ is large and negative, i.e., there is a strong attractive interaction between the chains. 
This is the expected result because, a greater proximity between the chains, would lead to a greater interaction between them. However, this fact also leads to an unexpected consequence. Indeed,  because in this case the dominant interaction takes place between dipoles belonging to different chains, the thermalization of the system is delayed because there is a sustained and significant energy flow between the two chains.

\section{Conclusions}
In this study we have explored the energy transfer dynamics in a one-dimensional ladder array of two chains of rigid rotating dipoles in mutual interaction. Periodic boundary conditions within each chain of the array have been assumed. We have focused on the planar dynamics of the system, that is, the dipoles are restricted to rotate in a common plane which is an invariant manifold of the array under the dynamics.

First, we have determined the ground state (GS) equilibrium configurations of the system and their energies as a function of the normalized distance $b=d/a$, being $d$ the separation between the two chains of the array, and $a$ the distance between two consecutive dipoles in a chain. We have found that there exists a critical value $b_c$ of the normalized distance, that separates two different GS equilibrium configurations. For values $b>b_c$, the GS, named $C_2$, is a head-tail configuration where all the dipoles are oriented along the array axis, but in opposite way in each chain. For values $b<b_c$, the GS, named $C_3$, is a configuration where all dipoles are oriented perpendicular to the array axis, in a head-tail configuration between both chains, alternating the orientation along the array axis.   

Prior to explore the energy transfer dynamics, we have determined up to what order we should be accounting for the interaction between neighboring dipoles in order to describe accurately the system behavior. To this end, we have studied the linear approximation of the array dynamics around the GS equilibrium configurations $C_2$ and $C_3$. In this context, we have deduced the expressions of the two bands of the dispersion relation as a function of the interaction order $r$, i.e. the order of the neighboring dipoles taken into account in the dipolar interaction. The analysis of the evolution with $r$ of the bands of the dispersion relation has shown that the inclusion of more dipoles beyond the interaction order $r=3$, does not alter significantly the linear behavior of the system around both GS configurations. Interaction order $r=3$ means that each dipole is coupled with its thirteen nearest neighboring dipoles.

According to above, we have extended the dipolar interaction up to order $r=3$ in the exploration of the energy propagation within the array of dipoles. The system, initially in the $C_2$ equilibrium configuration, is excited by supplying it with an excess energy $\Delta K$ to one of the dipoles. For these initial conditions, we have studied the time evolution of the array for different values of the system parameters $b \in [0.9,2.5]$ and $\Delta K \in [1,15]$. We have focused on two features of the energy flow: (i) We studied how the excess energy $\Delta K$ is shared between both chains of the array, that is, the energy stored in each chain; (ii) we analyzed how the energy is distributed within each chain, i.e. the localization of the energy along each chain. Consecuently, our analysis tools in the study of the energy transfer have been the time averaged normalized energy stored in each chain $\langle K_{u,v}\rangle$, and the time averaged participation functions $\langle \Pi_{u,v}\rangle$ which quantify the localization of the energy within each chain of the array. These time averaged quantities have been determined in three convenient time intervals $\Delta t$ distributed along all the computation time, which allow us to analyze the energy transfer process from early to late time instants.

With regard to the energy distribution between both chains of the array, we have found that in most cases from the early stage of the time evolution, the energy of the array is evenly distributed between both chains, that is, $\langle K_u\rangle = \langle K_v\rangle = 0.5$. Only for the parameter values $\Delta K \gtrsim 7$ and $b \gtrsim 1.5$, the system energy is largely stored in the active chain, that is, the chain initially excited. As expected, for intermediate and late times, this region of non-evenly energy sharing between both chains, suffers a progressive reduction in size. Nevertheless, it is worth to note that, even up to the maximum computation time used in this study, there still persists a region $(b\gtrsim 2$ and $7 \lesssim \Delta K \lesssim 9)$ where the energy equipartition between the two chains has not been achieved.

With respect to the energy distribution along each chain, we have found that, when the system reaches the thermal equilibrium, the participation functions of each chain attain the maximum value $\langle \Pi_u\rangle=\langle \Pi_v\rangle=\langle \Pi_{eq}\rangle\approx 0.6$. This value is very close to the corresponding equilibrium value for a single linear chain of dipoles \cite{chain}. This thermal equilibrium value for the participation functions is reached in both chains for early times  for most of the values of the parameters $b$ and $\Delta K$. It is worth to highlight that, for the active chain, there is a region in the parameter space $(b,\Delta K)$ where the thermalization is not achieved even up to the maximum computation time because the energy remains strongly localized within that chain. The size and time evolution of this region match very well with the region of non-equipartition energy between both chains. The similar time evolution of these two energy features can be explained by the existence of persistent chaotic breathers that propagate in the active chain, where a significant amount of energy is localized in a few neighboring dipoles. The presence of these chaotic breathers prevents both the energy equipartition between chains, and the thermal equilibrium within the active chain. The thermalization is not reached in both chains for low values of the system parameters $b$ and $\Delta K$ despite both chains are in mutual energy equipartition. This behavior can be explained by analyzing the evolution of the energy involved in the mutual interaction between both chains as a function of the normalized distance $b$. When the separation between the two chains is small (low values of $b$), we have found that the mutual interaction energy is large and negative, which means a strong and attractive interaction between the chains. Therefore, the dominant interaction in the array is between dipoles of different chains, and the thermal equilibrium within each chains is delayed because the main energy flow takes place between the two chains of the array. 

In this paper, we have restricted ourselves to only an invariant subspace of the system dynamics. A natural continuation would be the study of the energy transfer in the full dimensional dynamics of the same array. A next step could be to consider more complex configurations of one-dimensional arrays of dipoles such as diamond or saw-tooth like arrays \cite{A1021}, or even dimerized versions of them. In this sense, an interesting line would be to study the existence of flat bands \cite{A1022,A1160,A1161} in the dispersion relations of this kind of one-dimensional arrays, and their effect on the energy transfer dynamics.

\section{Acknowledgments}
M.I. and J.P.S. acknowledge financial support by the Spanish Proyect PID2022-140469NB-C22 (MICIN). This work used the Beronia cluster (Universidad de La Rioja), which is supported by FEDER-MINECO Grant No. UNLR-094E-2C-225.
R.G.F. gratefully acknowledges financial support by the Spanish project PID2020-113390GB-I00 (MICIN), and the Andalusian research group FQM-207.

\appendix
\section{$C_2$ equilibrium configuration: Linear behavior in the neighborhood of the center and boundaries of the Brillouin zone}
\begin{figure}[h]
\centerline{\includegraphics[scale=0.4]{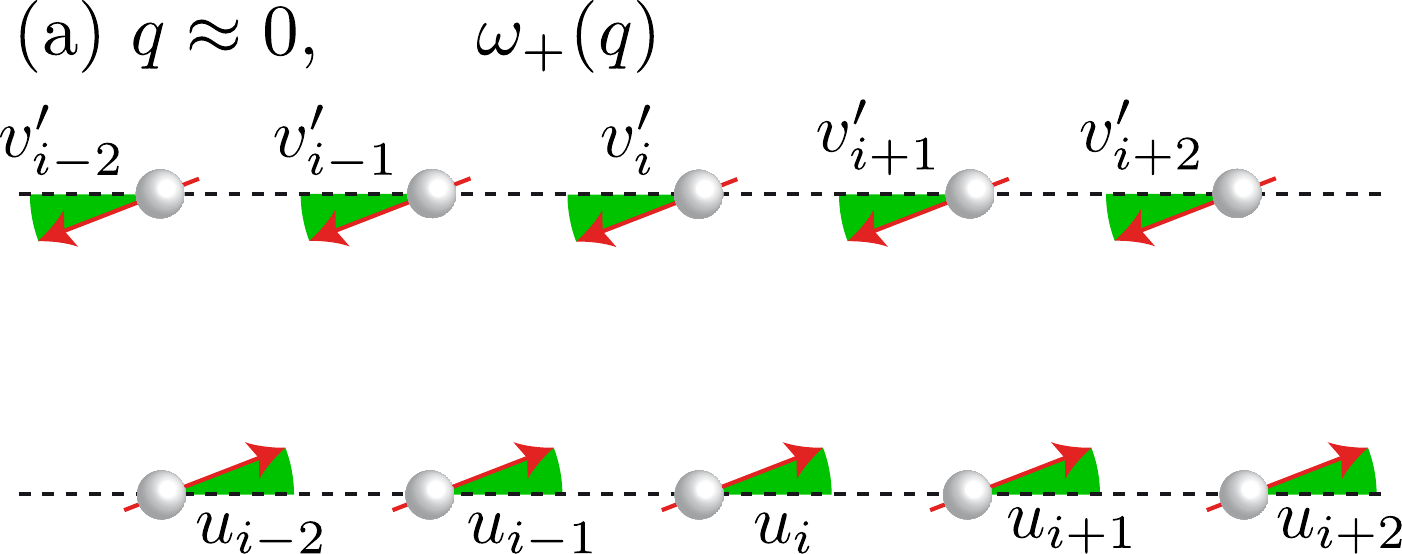}}
\vspace{1cm}
\centerline{\includegraphics[scale=0.4]{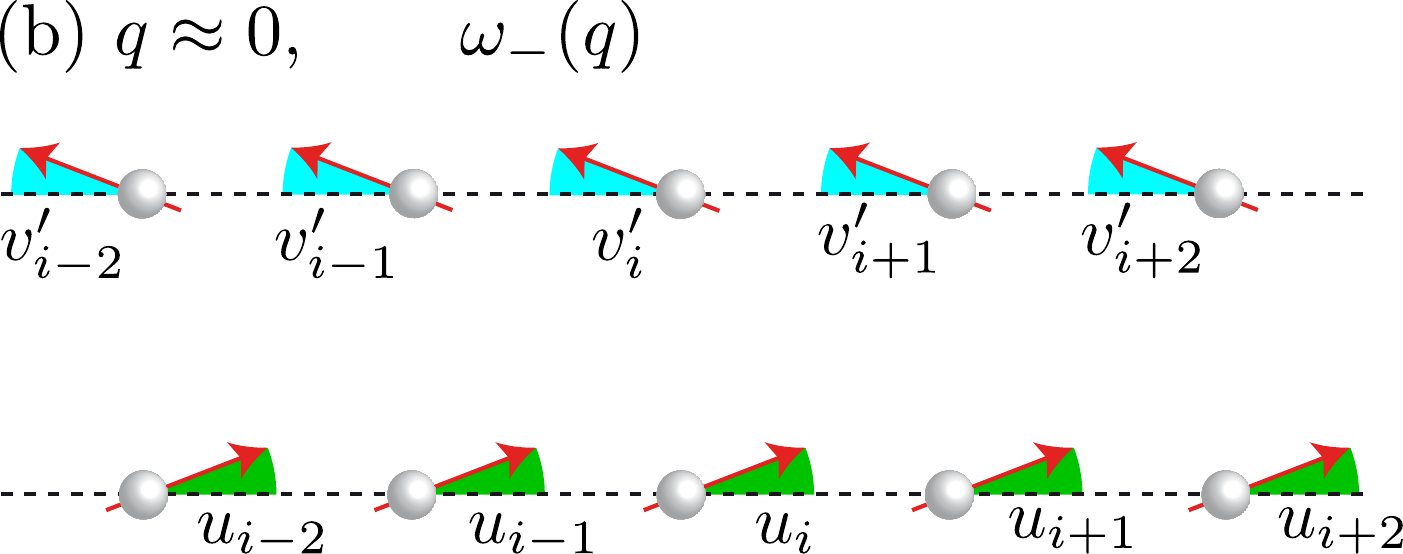}}
\vspace{1cm}
\centerline{\includegraphics[scale=0.4]{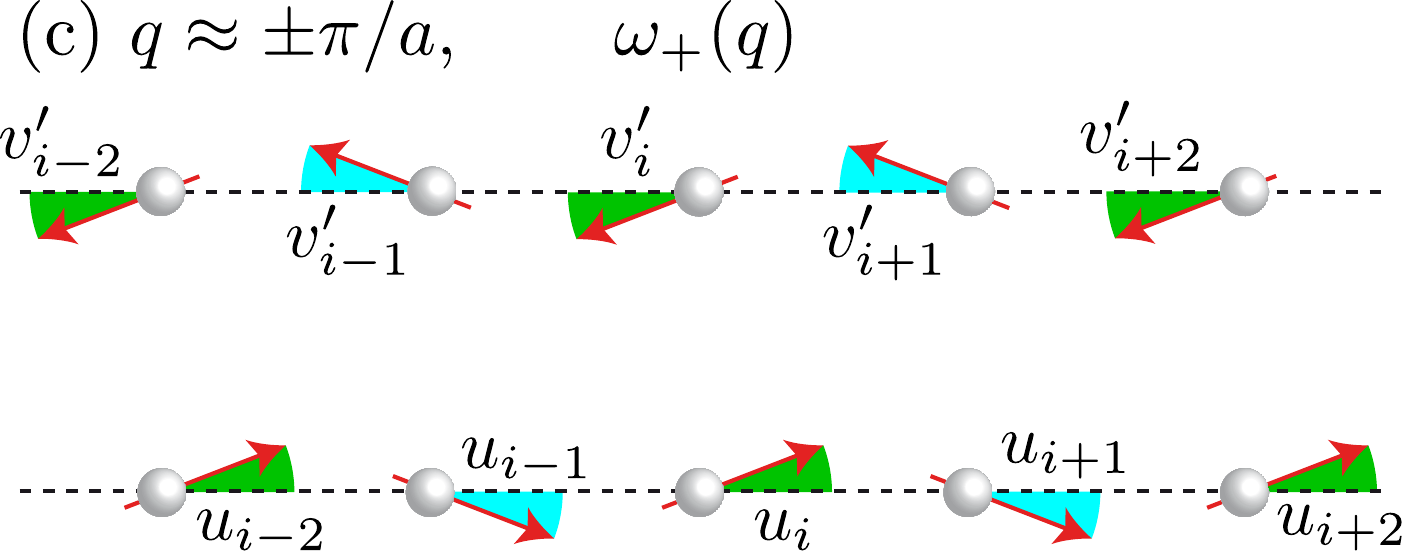}}
\vspace{1cm}
\centerline{\includegraphics[scale=0.4]{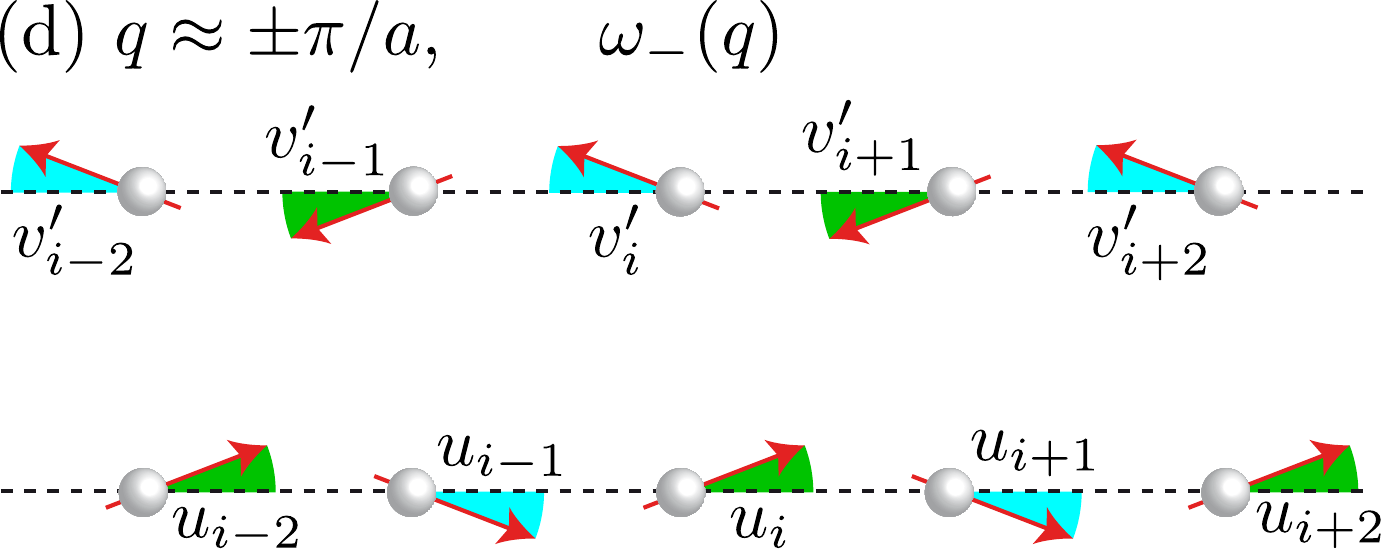}}
\caption{Linear oscillations of the dipoles in the $\omega_\pm$ dispersion bands in the neighborhoods of the center $(q=0)$ and boundaries $(q=\pm \pi/a)$ of the Brillouin zone in the $C_2$ equilibrium configuration. Green color stands for positive oscillation angles and blue color for negative oscillation angles.}
\label{fi:linearoscillationsC2}
\end{figure}

Around the $C_2$ equilibrium, the ratio of the amplitudes $A(q)$ and $B(q)$ of the propagating plane waves \ref{waves} is determined by Eq.\ref{ratioAB}. In the neighborhood of the center of the Brillouin zone $(q \approx 0)$, we have that the plane waves \ref{waves} take the expressions
\begin{equation}
u_n(t) \approx A(q) \exp^{-i \omega t}, \qquad v_n'(t) \approx B(q) \exp^{-i \omega t}.
\end{equation}
Then, around the center of the dispersion bands of Eq.\ref{omega}, within each chain of the array, all dipoles oscillate in phase. Besides, taking into account the ratio \ref{ratioAB} between the amplitudes, in the positive band $\w_+$ all dipoles of the ladder chain oscillate in phase, so that $u_n(t)-v_n'(t)=0$, i.e. $u_n(t)-v_n(t)=\pi$ (see Fig. \ref{fi:linearoscillationsC2}(a)). On the contrary, in the negative band $\w_-$ of \ref{omega} the dipoles of the upper chain of the array oscillate in opposite phase with respect to the dipoles of the lower chain, that is, $u_n(t)+v_n'(t)=0$, i.e. $u_n(t)+v_n(t)=\pi$  (see Fig. \ref{fi:linearoscillationsC2}(b)).
On the other hand, at the boundaries of the Brillouin zone $(q \approx \pm \pi/a)$, the propagating waves \ref{waves} can be written as
\begin{equation}
u_n(t) \approx (-1)^n A \exp^{-i \omega t}, \qquad v_n'(t) \approx (-1)^n B \exp^{-i \omega t}.
\end{equation}
Therefore, around the ends of dispersion bands \ref{omega}, within each chain, the nearest neighbor dipoles oscillate opposite in phase. Moreover, taking into account the ratio \ref{ratioAB}, in the positive band $\w_+$ of \ref{omega} the pair of dipoles located in the same position in each chain oscillate in phase (see Fig.\ref{fi:linearoscillationsC2}(c)), while in the negative band $\w_-$ 
of \ref{omega} they oscillate in opposite phase (see Fig. \ref{fi:linearoscillationsC2}(d)).

\section{$C_3$ equilibrium configuration: Linear behavior in the neighborhood of the center and boundaries of the Brillouin zone}
\begin{figure}[h]
\centerline{\includegraphics[scale=0.4]{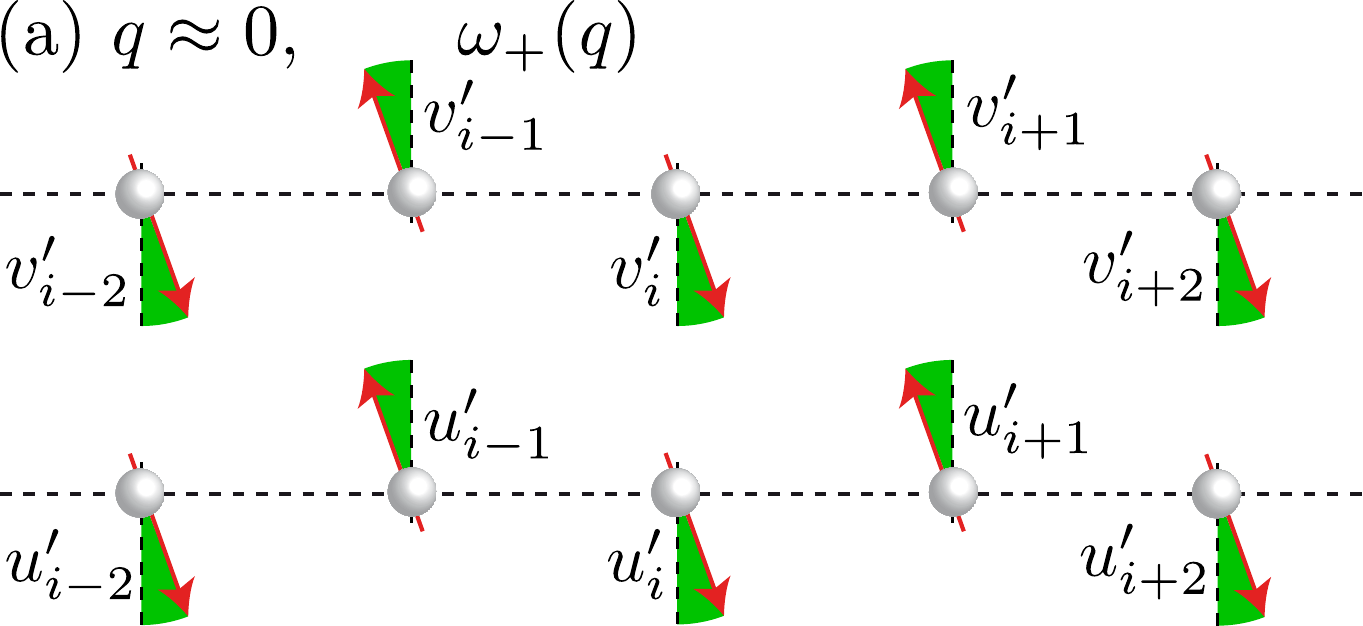}}
\vspace{1cm}
\centerline{\includegraphics[scale=0.4]{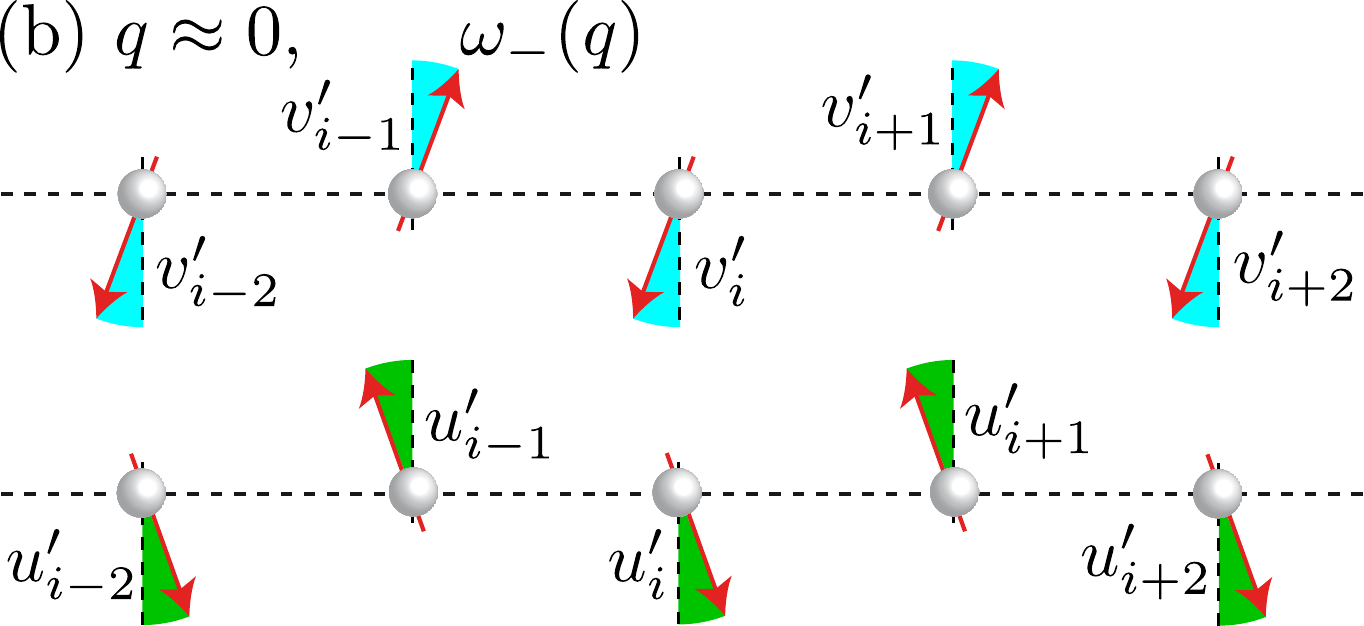}}
\vspace{1cm}
\centerline{\includegraphics[scale=0.4]{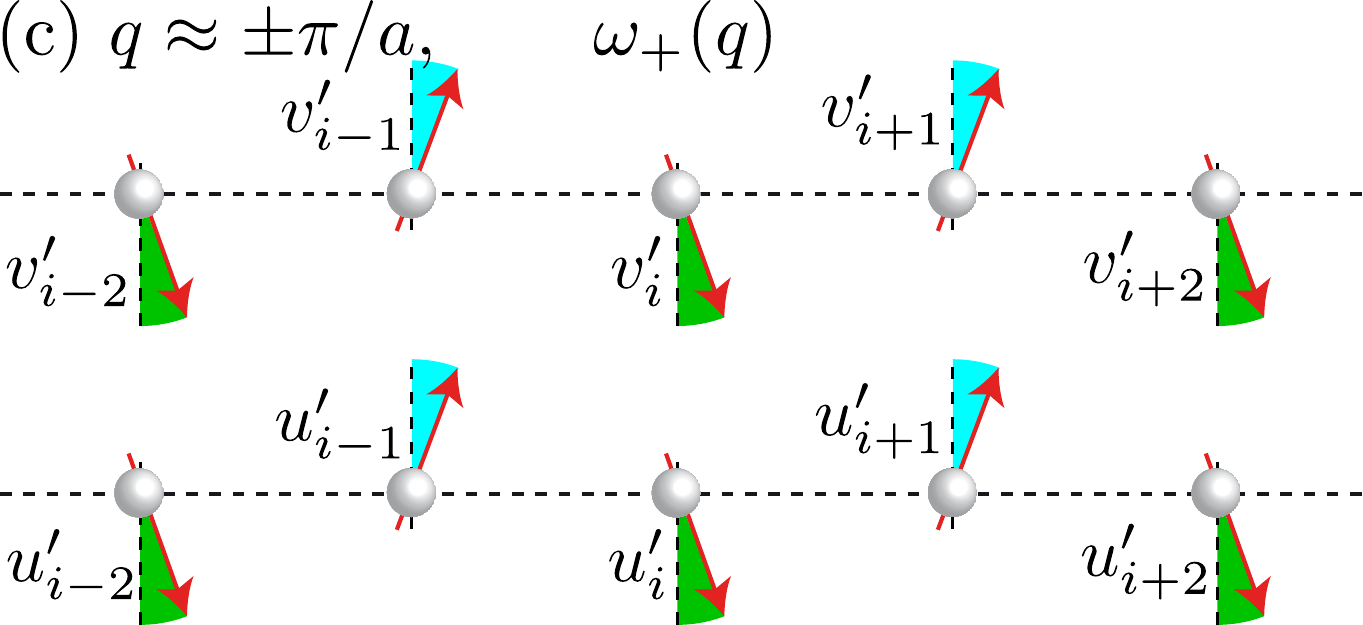}}
\vspace{1cm}
\centerline{\includegraphics[scale=0.4]{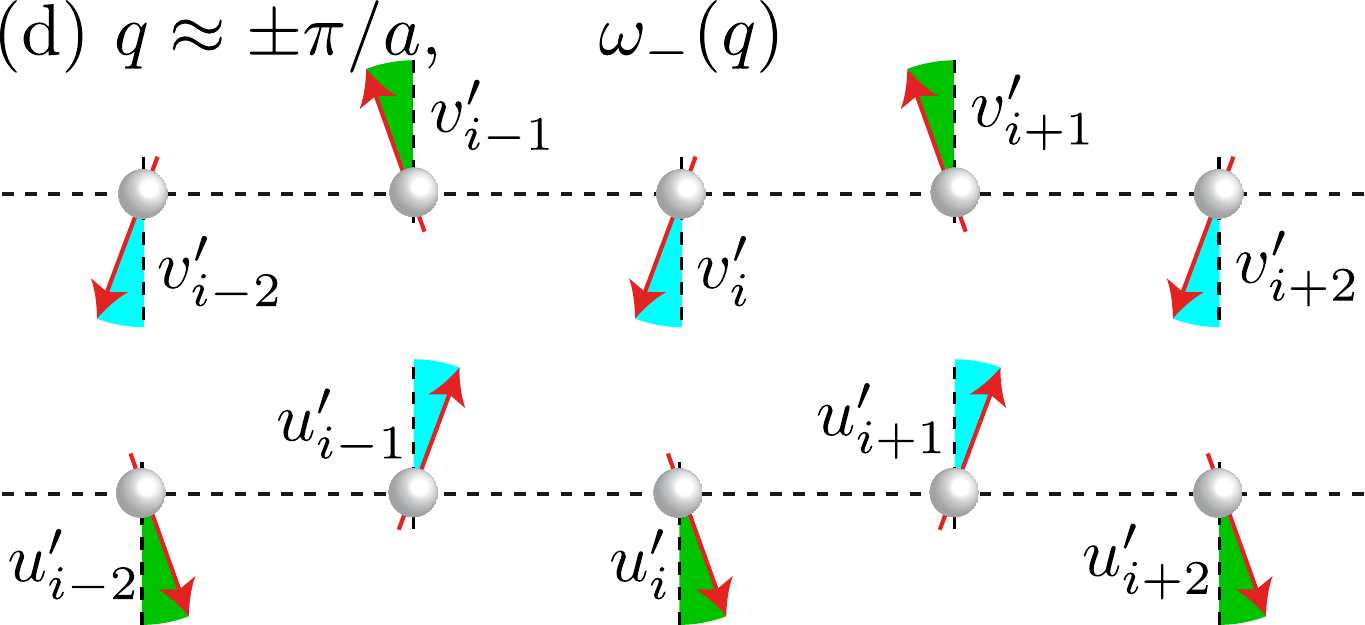}}
\caption{Linear oscillations of the dipoles in the $\omega_\pm$ dispersion bands in the neighborhoods of the center $(q=0)$ and boundaries $(q=\pm \pi/a)$ of the Brillouin zone in the $C_3$ equilibrium configuration. Green color stands for positive oscillation angles and blue color for negative oscillation angles.}
\label{fi:linearoscillationsC3}
\end{figure}

Around the $C_3$ equilibrium, the ratio of the amplitudes $A(q)$ and $B(q)$ of the propagating plane waves \ref{waves} is also determined by Eq.\ref{ratioAB}.
In the neighborhood of the center of the Brillouin zone $(q=0)$, in the dispersion bands $\w_{\pm}$ (see Eq.\ref{omega2}, within each chain of the array, all dipoles oscillate in phase. Besides, taking into account the ratio \ref{ratioAB} between the amplitudes, in the positive band $\w_+$ all dipoles of the ladder chain oscillate in phase, so that $u_n'(t)-v_n'(t)=0$ (see Fig. \ref{fi:linearoscillationsC3}(a)), whereas in the negative band $\w_-$ the dipoles of the upper chain oscillate in opposite phase with respect to the dipoles of the lower chain, that is $u_n'(t)+v_n'(t)=0$ (see Fig. \ref{fi:linearoscillationsC3}(b)).
On the other hand, at the boundaries of the Brillouin zone $(q=\pm \pi/a)$, within each chain, the nearest neighbor dipoles oscillate in opposite phase, so that $u_n'(t)+u_{n+i}'(t)=0$ and $v_n'(t)+v_{n+i}'(t)=0$. Taking into account the ratio \ref{ratioAB}, in the positive band $\w_+$ the pair of dipoles located in the same position in each chain oscillate in phase (see Fig.\ref{fi:linearoscillationsC3}(c)), while in the negative band $\w_-$ they oscillate with opposite phase (see Fig. \ref{fi:linearoscillationsC3}(d)).

\end{document}